\documentclass[journal,12pt,onecolumn,draftclsnofoot]{IEEEtran}

\usepackage{cite}
\usepackage{hyperref}

\ifCLASSINFOpdf
\usepackage[pdftex]{graphicx}
\else
\fi

\usepackage{bm}
\usepackage{dsfont}
\usepackage{amsthm, amsmath, amssymb}
\usepackage{mathrsfs}
\usepackage{subeqnarray}
\usepackage{cases}

\usepackage{algorithm}  
\usepackage{algpseudocode}  
\usepackage{amsmath}

\usepackage{array}
\usepackage{booktabs}

\usepackage{subfigure}

\begin{document}

\bibliographystyle{IEEEtran}

%
\title{Network Topology Inference Based on Timing Meta-Data}

%
%

\author{Wenbo Du,~\IEEEmembership{Member,~IEEE},
        Tao Tan,
        Haijun Zhang,~\IEEEmembership{Senior Member,~IEEE}, \\
        Xianbin Cao,~\IEEEmembership{Senior Member,~IEEE},
        Gang Yan,~\IEEEmembership{Member,~IEEE}, \\
        and Osvaldo Simeone,~\IEEEmembership{Fellow,~IEEE}
\thanks{This work was supported in part by the National Key Research and Development Program of China under Grant 2019YFF0301400, in part by the National Natural Science Foundation of China under Grant 61961146005, in part by the Shuohuang Railway Project under Grant GJNY-19-90. The work of O. Simeone was supported by the European Research Council (ERC) under the European Union’s Horizon 2020 Research and Innovation Programme (Grant Agreement No. 725731) and by an EPSRC Open Fellowship.
}
\thanks{W. Du, T. Tan and X. Cao are with the School of Electronic and Information Engineering, Beihang University, Beijing 100191, China, with the Key Laboratory of Advanced Technology of Near Space Information System (Beihang University). (e-mail: wenbodu@buaa.edu.cn; tantao@buaa.edu.cn; xbcao@buaa.edu.cn).}
\thanks{H. Zhang is with Beijing Engineering and Technology Research Center for Convergence Networks and Ubiquitous Services, University of Science and Technology Beijing, Beijing, China, 100083 (e-mail: haijunzhang@ieee.org).}
\thanks{G. Yan is with School of Physics Science and Engineering, Tongji University, Shanghai 200092, China (e-mail: eegyan@gmail.com).}
\thanks{O. Simeone is with the King’s Communications, Learning, and Information Processing (KCLIP) Laboratory, Department of Engineering, King’s College London, London WC2R 2LS, U.K. (e-mail: osvaldo.simeone@kcl.ac.uk).}
}

\markboth{}
{Shell \MakeLowercase{\textit{et al.}}: Network Topology Inference based on Timing Meta-Data}

\maketitle

\begin{abstract}

Consider a processor having access only to meta-data consisting of the timings of data packets and acknowledgment (ACK) packets from all nodes in a network. The meta-data report the source node of each packet, but not the destination nodes or the contents of the packets. The goal of the processor is to infer the network topology based solely on such information. Prior work leveraged causality metrics to identify which links are active. If the data timings and ACK timings of two nodes -- say node 1 and node 2, respectively -- are causally related, this may be taken as evidence that node 1 is communicating to node 2 (which sends back ACK packets to node 1). This paper starts with the observation that packet losses can weaken the causality relationship between data and ACK timing streams. To obviate this problem, a new Expectation Maximization (EM)-based algorithm is introduced -- EM-causality discovery algorithm (EM-CDA)  -- which treats packet losses as latent variables. EM-CDA iterates between the estimation of packet losses and the evaluation of causality metrics.  The method is validated through extensive experiments in wireless sensor networks on the NS-3 simulation platform.

\end{abstract}
\begin{IEEEkeywords}
Network topology inference, meta-data, causality metrics, packet loss, expectation maximization.
\end{IEEEkeywords}
%
\IEEEpeerreviewmaketitle

\section{Introduction} \label{Introduction}

\subsection{Motivation and Overview}

\begin{figure}[htp]
    \centering
    \includegraphics[scale=0.32]{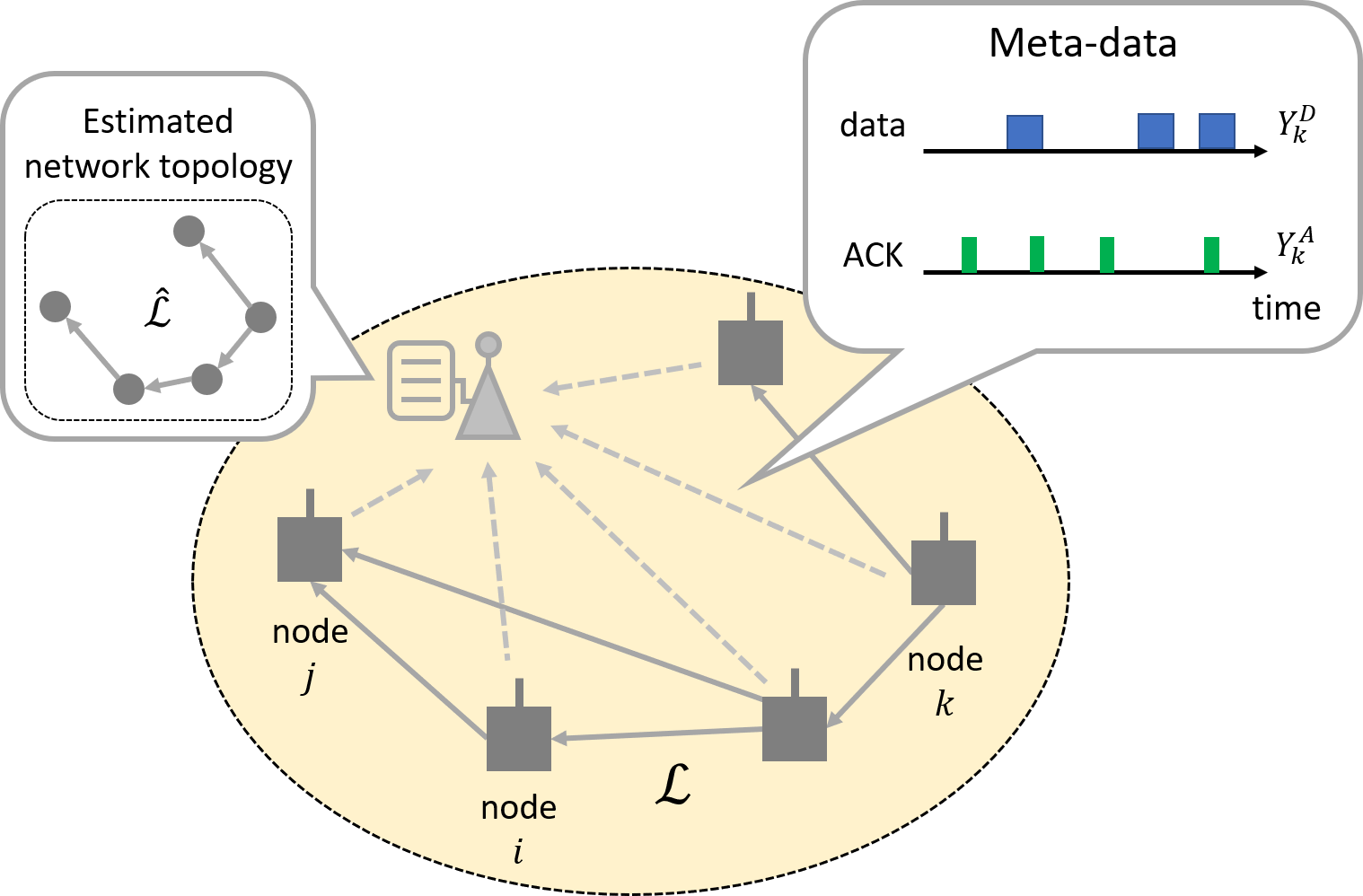}
    \caption{An example of a wireless device-to-device network with a set of nodes $\mathcal{N} = \{1, 2, 3, 4, 5 \}$ and a set of directional links $\mathcal{L} = \{ (2, 1), (3, 1), (3, 2), (4, 3), (4, 5) \}$. The central monitor collects meta-data from the nodes in the form of data and acknowledgment (ACK) packet timings, based on which it aims to estimate the network topology. Unlike previous work \cite{laghate2017learning, sharma2019communication}, this paper allows for packet losses, making it more challenging to interpret and use meta-data.}
    \label{fig-model}
\end{figure}

Information about the topology of a device-to-device wireless network, e.g., a sensor network, is essential to implement functionalities such as routing, anomaly detection, and load balance. In recent years, \emph{passive} monitoring methods that leverage only observations of network traffic have received significant attention, owing to their cost-effectiveness as compared to \emph{active} methods that probe nodes for information \cite{cociglio2019multipoint, brissaud2019transparent}. Passive monitoring methods can be ``invasive'',  implementing packet inspection techniques like demodulation and decryption \cite{gao2014ipath}; or ``non-invasive'', leveraging only meta-data. Invasive methods can achieve high accuracy, but they require complex sensors and baseband processors. Non-invasive techniques have the advantage of requiring only information about the timings of data packet and acknowledgement (ACK) packets, which is relatively easier to collect and process (see Fig. \ref{fig-model}).  This paper contributes to the line of work on passive, non-invasive, network topology estimation.

To elaborate, consider, as in Fig. \ref{fig-model}, a processor having access only to meta-data consisting of the timings of data packets and ACK packets from all nodes in a network. The meta-data report the source node of each packet, but not the destination nodes or the contents of the packets. The goal of the processor is to infer the network topology based solely on such information. Reference \cite{tilghman2013inferring} proposed to leverage causality metrics to identify which links are active. The key underlying idea is that, if the data timings and ACK timings of two nodes -- say node 1 and node 2, respectively -- are causally related, this may be taken as evidence that node 1 is communicating to node 2 (which sends back ACK packets to node 1). The same principle underpins network discovery in fields as diverse as biology and sociology \cite{guo2021multiscale, finkle2018windowed, runge2019inferring, bovet2019influence}.

The causality discovery algorithm (CDA) introduced in \cite{tilghman2013inferring} was based on Granger causality, a measure of causal dependence based on auto-regressive modelling \cite{granger1969investigating}. Asymmetric Granger causality was used in \cite{laghate2017learning}, which outperforms GCT at a finer time resolution. Transfer Entropy (TE) was then adopted for CDA in \cite{sharma2019communication}. TE has the advantage of capturing also non-linear causality relationships \cite{schreiber2000measuring, staniek2008symbolic, zhang2019itene}.

This paper starts with the observation that packet losses can weaken the causality relationship between data and ACK timing streams. To obviate this problem, a new Expectation Maximization (EM)-based algorithm is introduced -- EM-causality discovery algorithm (EM-CDA)  -- which treats packet losses as latent variables.

\subsection{Related Work}

Active probing is a traditional method used in wireless topology inference, whereby information is collected from neighboring nodes \cite{liang2017cooperative, liu2020cooperative}. In such methods, a subset of “privileged” nodes usually performs the probing task \cite{deb2002topology}. While these schemes can potentially infer accurately the functional network without location information, the energy cost associated with active methods is a critical drawback.

As for passive schemes, references \cite{zeng2008blindly, partridge2002using} exploit spectral coherence to infer the network topology, but this approach tends to detect spurious links. In \cite{moore2016analysis, xu2016learning}, multivariate Hawkes processes, a parametric formulation of packet arrival statistics, is considered to recover the network topology. These solutions are model-based, and hence operate under strict assumptions on the valid of the model. CDA-based passive topology inference methods currently provide state-of-the-art results for passive topology inference. Apart from the papers reviewed in the previous subsection, the authors of \cite{testi2019machine} leverage blind source separation to improve the problem caused by interference. The work \cite{elsegai2019granger} considers an equidistant missing-data problem based on Granger causality. Nonetheless, the problem of missing observations caused by packet loss is still an open issue, which can result in a significant drop in inference accuracy \cite{tilghman2013inferring, laghate2017learning, sharma2019communication, testi2020blind}.

\subsection{Main Contributions}

Addressing the need for passive topology inference techniques that are robust to packet losses, this paper introduces EM-CDA. The main contributions of this paper can be summarized as follows.

\begin{enumerate}
    \item[$\bullet$] We formulate the problem of network topology inference as the maximum likelihood problem of estimating existing network links in the presence of latent variables representing packet losses. EM-CDA is derived as a tractable approximation of the resulting EM algorithm. Accordingly, EM-CDA iterates between the estimation of packet losses and the evaluation of causality metrics based on the estimated missing packets.
    \item[$\bullet$] EM-CDA is validated through experiments in the wireless network on the NS-3 simulation platform, demonstrating that EM-CDA can improve the detection probability and false alarm probability rate of CDA ranging from 4\% to 12\% under a variety of practical conditions.
\end{enumerate}

The rest of this paper is organized as follows. The wireless network scenario and system model are described in Section \ref{System Model and Problem Formulation}. The state-of-the-art causality discovery algorithm (CDA) for wireless network topology inference is presented in Section \ref{Causality method}. EM-CDA scheme is introduced in Section \ref{EM formulation}. In Section \ref{Numerical Results}, numerical results are given to demonstrate the performance of the proposed algorithm. Finally, the paper is concluded in Section \ref{Conclusion}.


\section{System Model and Problem Setup} \label{System Model and Problem Formulation}

In this section, we describe the setting under study in which, as illustrated in Fig. \ref{fig-model}, a central monitor collects meta-data about packet timings from the nodes of a network in order to infer the network topology.
In this paper, unlike \cite{laghate2017learning, sharma2019communication}, we allow packet losses to occur on the communication links. This creates additional challenges in relating the timings of data and control (acknowledgment) packets, motivating the novel estimation algorithm introduced in the next section.

\subsection{Setting}

Consider the problem of estimating the topology of a network consisting of a set $\mathcal{N}=\{1, 2, ..., N\}$ of $N$ nodes and of a set $\mathcal{L}=\left\{ (i, j) \big| i, j \in \mathcal{N} \right\}$ of $M \leq N(N-1)$ \textit{directional} links. The presence of a link $(i, j) \in \mathcal{L}$ with $i, j \in \mathcal{N}$ and $i \neq j$ indicates that node $i$ communicates with node $j$.

As in \cite{laghate2017learning, sharma2019communication}, we assume that a central monitor collects meta-data in the form of transmission timestamps reporting the time instants at which data packets or acknowledgments (ACKs) are sent by each node within a given time window.
Only timing meta-data is collected, and hence the monitor is only aware of packet timings, and not of the intended destination of any given packet.
Successful transmission of a data packet from one node to another causes the transmission of an ACK from the receiving node to the transmitting one. ACKs are assumed to be much shorter than data packets and not subject to data losses.

\subsection{Data Transmission and Channel Model}

The observation period $T$ is discretized into $K$ equal time slots of duration $T_s=T/K$, which are indexed by integer $k \in \mathcal{K} = \{1, 2, ..., K\}$. To describe the timing information recorded by node $i \in \mathcal{N}$, two integer-valued time sequences $Y_{i}^{D}[k]$ and $Y_{i}^{A}[k]$ are introduced, corresponding to data packets and ACKs, respectively. The data packet timing sample $Y_{i}^{D}[k]$ equals the number of data packets sent by node $i$ in time slot $k$. In a similar way, the timing information sample $Y_i^A[k]$ for ACK packets equals the number of ACK packets sent by node $i$ in time slot $k$.
We collect the data timing information across all time slots for node $i$ in the $K \times 1$ vector
\begin{equation} \label{per-node-d}
    \mathbf{Y}_{i}^{D} = \left[ Y_{i}^{D}[1], Y_{i}^{D}[2], ..., Y_{i}^{D}[K] \right]^{\mathsf{T}},
\end{equation}
and the ACK timing information in the $K \times 1$ vector
\begin{equation} \label{per-node-a}
    \mathbf{Y}_{i}^{A} = \left[ Y_{i}^{A}[1], Y_{i}^{A}[2], ..., Y_{i}^{A}[K] \right]^{\mathsf{T}}.
\end{equation}

The data and ACK timing series for node $i$ can be expressed as the sum of individual contributions corresponding to the distinct communication links stemming from node $i$. To elaborate, we define the per-link binary sequences
\begin{equation} \label{per-link-d}
    Y_{i, j}^{D}[k] = 
    \left\{ \begin{array}{ll}
    1 & \mbox{if a data packet is sent on link $(i,j)$ in time slot $k$}, \\
    0 & \mbox{otherwise},
    \end{array}\right.
\end{equation}
and
\begin{equation} \label{per-link-a}
    Y_{i, j}^{A}[k] = 
    \left\{ \begin{array}{ll}
    1 & \mbox{if an ACK is sent on link $(i,j)$ in time slot $k$}, \\
    0 & \mbox{otherwise}.
    \end{array}\right.
\end{equation}
Note that the time slot $T_s$ is assumed to be sufficiently small so that no more than one data packet is sent by a node to another node within a single slot.
Using the per-link sequences (\ref{per-link-d})-(\ref{per-link-a}), the per-node observations (\ref{per-node-d})-(\ref{per-node-a}) can be written as the sums
\begin{equation} \label{DATA}
    Y_i^{D}[k] = \sum_{(i, j) \in \mathcal{L}}Y_{i, j}^{D}[k],
\end{equation}
and
\begin{equation} \label{ACK}
    Y_i^{A}[k] = \sum_{(i, j) \in \mathcal{L}}Y_{i, j}^{A}[k].
\end{equation}
Importantly, by collecting the sequences (\ref{DATA})-(\ref{ACK}), the monitor only has aggregate information regarding the achieving of each node $i$ while not having access to the per-link series $Y_{i, j}^{D}[k]$ and $Y_{i, j}^{A}[k]$.

The data packet and ACK timing sequences are related by the ARQ protocol. Let us denote as $\tau_{i,j}[k]$ the delay, measured in the number of time slots, between the transmission of a data packet in time slot $k$ by node $i$ to node $j$ and the transmission of the corresponding ACK packets from node $j$ to node $i$. We also introduce the per-link binary error variable $E_{i, j}[k]$ defined as
\begin{equation} \label{error}
    E_{i, j}[k] = 
    \left\{ \begin{array}{ll}
    1 & \mbox{if an error occurs on link $(i, j)$ in time slot $k$}, \\
    0 & \mbox{otherwise}.
    \end{array}\right.
\end{equation}
With these definitions, we have the equality
\begin{equation} \label{receive-one}
    Y_{j, i}^{A}[k + \tau_{i, j}[k]] = \left(1 - E_{i, j}[k] \right) Y_{i, j}^{D}[k],
\end{equation}
which indicates that an ACK is sent in time slot $k + \tau_{i, j}[k]$ on link $(j, i)$, i.e., $Y_{j, i}^{A}[k + \tau_{i, j}[k]] = 1$, when a data packet is sent in time slot $k$ on the reverse link $(i, j)$, i.e., $Y_{i, j}^{D}[k]=1$ and an error does not occur on link $(i, j)$, i.e., $E_{i, j}[k]=0$. The $\tau_{i, j}[k]$ may severely vary across links and time slots, and it is unknown to the monitor.

\subsection{Topology Inference}

The timing information sequences $\left\{ \mathbf{Y}_{i}^{D} \big| \forall i \in \mathcal{N} \right\}$ and $\left\{ \mathbf{Y}_{i}^A \big| \forall i \in \mathcal{N} \right\}$ in (\ref{per-node-d})-(\ref{per-node-a}) collected from all nodes are used by the monitor to infer the topology, which is defined by set of links $\mathcal{L}$. 
The links set $\mathcal{L}$ can be equivalently also described by the adjacency matrix $\mathbf{A} = \left\{a_{i, j} \big| \forall i, j \in \mathcal{N} \right\}$ with entries
\begin{equation}
    a_{i, j} =
    \left\{ \begin{array}{ll}
    1 & \mbox{if $(i, j) \in \mathcal{L}$}, \\
    0 & \mbox{otherwise}. 
    \end{array}\right.
\end{equation}
Therefore, the goal of the monitor is to use sequence $\left\{ \mathbf{Y}_{i}^{D} \big| \forall i \in \mathcal{N} \right\}$ and $\left\{ \mathbf{Y}_{i}^A \big| \forall i \in \mathcal{N} \right\}$ to produce an estimate $\hat{\mathbf{A}}$ of the adjacency matrix $\mathbf{A}$, or equivalently an estimate $\hat{\mathcal{L}}$ of the link set $\mathcal{L}$.

\section{Causality-based Topology Estimation} \label{Causality method}

In this section, we review the Causality Discovery Algorithms (CDAs) introduced in \cite{tilghman2013inferring, laghate2017learning, testi2020blind, sharma2019communication} wherein links are included in the estimated set $
\hat{\mathcal{L}}$ based on measures of causal dependence between data and ACK sequences of two nodes.

\subsection{Causality Discovery Algorithm}
In CDA schemes, the monitor estimates a measure of causal dependence 
$\Phi \left(\mathbf{Y}_i^D \rightarrow \mathbf{Y}_j^A \right)$ between sequences $\mathbf{Y}_i^D$ and $\mathbf{Y}_j^A$ for each pair of nodes $i$ and $j$.
The measure $\Phi \left(\mathbf{Y}_i^D \rightarrow \mathbf{Y}_j^A \right)$ quantifies the degree to which the future of sequence $\mathbf{Y}_j^A$ can be predicted based on the past of sequence $\mathbf{Y}_i^D$.
A link $(i, j)$ is added to the estimated set $\hat{\mathcal{L}}$ if the measure $\Phi \left(\mathbf{Y}_i^D \rightarrow \mathbf{Y}_j^A \right)$ is larger than some threshold $\theta_{i, j}$. This condition can be equivalently expressed as
\begin{equation} \label{inference form}
    \hat{a}_{i, j} = 
    \left\{ \begin{array}{ll}
    1 & \mbox{if $\Phi  \left( \mathbf{Y}_i^D \rightarrow \mathbf{Y}_j^A \right) > \theta_{i, j}$}, \\
    0 & \mbox{otherwise}. 
    \end{array}\right.
\end{equation}
The rationale for this decision rule is that, if link $(i, j)$ exists, then by (\ref{receive-one}) data packets from node $i$ cause ACKs from node $j$, assuming that there are no errors.
This, in turn, ideally contributes to increasing the causal dependence measure $\Phi \left(\mathbf{Y}_i^D \rightarrow \mathbf{Y}_j^A \right)$.

We now discuss specific choice for the causal dependence measure $\Phi \left(\mathbf{Y}_i^D \rightarrow \mathbf{Y}_j^A \right)$.

\subsection{Causality Metrics} \label{Causality Discovery}
Granger causality (GC) is a standard measure of causal dependence that is based on linear prediction. Given two time sequences $\mathbf{Y}_i^D$ and $\mathbf{Y}_j^A$, GC evaluates the extent to which omitting the past of time series $Y_i^D[k]$ increases the prediction error for sequence $Y_j^A[k]$ when prediction is based on a linear $R$-order autoregressive (AR) model.
Formally, GC uses the available observations $\mathbf{Y}_i^D$ and $\mathbf{Y}_j^A$ to fit separately two models, namely
\begin{equation} \label{GCT-H1}
    Y_j^A[k] = \sum_{r=1}^R a_{1r}Y_j^A[k-r] + \sum_{r=1}^R a_{2r}Y_i^D[k-r] + \varepsilon_{k},
\end{equation}
and
\begin{equation} \label{GCT-H0}
    Y_j^A[k] = \sum_{r=1}^R b_{r}Y_j^A[k-r] + \eta_{k},
\end{equation}
by optimising over parameters ${\left\{ a_{1r}, a_{2r} \right\}}_{r=1}^R$, and $\{{b_{r}\}}_{r=1}^R$
via least squares minimization.
In (\ref{GCT-H1})-(\ref{GCT-H0}), the quantities $\varepsilon_{k}$ and $\eta_{k}$ represent the prediction residuals.
The prediction residuals $\varepsilon_{k}$ in (\ref{GCT-H1}) account for prediction errors accrued on the ACK sequence $Y_j^A[k]$ when the past of data packet sequence $Y_i^D[k]$ is known; while the residuals $\eta_{k}$ in (\ref{GCT-H0}) are obtained when prediction can only use the past sample for the ACK sequence $Y_j^A[k]$ itself.
The GC-based measure is given by \cite{tilghman2013inferring}
\begin{align} \label{GCT}
    \Phi_{\rm GC} \left(\mathbf{Y}_i^D \rightarrow \mathbf{Y}_j^A \right) = \frac{(\sum_{k=1}^H |\eta_{k}|^2 - \sum_{k=1}^H |\varepsilon_{k}|^2) / R}{\sum_{k=1}^H |\varepsilon_{k}|^2 / (K-3R-1)},
\end{align}
which is large when the sum-residual $\sum_{k=1}^H |\eta_{k}|^2$ is larger than $\sum_{k=1}^H |\varepsilon_{k}|^2$, where $H=K-R$.
GC was used in \cite{tilghman2013inferring, testi2020blind} for topology estimation.

Transfer entropy (TE) is an information-theoretic causality measure that does not assume a linear relation between sequences $\mathbf{Y}_i^D$ and $\mathbf{Y}_j^A$ as GC.
To introduce it, let us define as $I\left(A; B \big| C \right)$ the conditional mutual information of random variable $A$ and $B$ given $C$, which is defined as
\begin{equation}
    I \left( A;B \big| C \right)
    = \mathbb{E} \left[ \log_2 \frac{p\left( A \big| B, C \right)}{p\left( A \big| C \right)} \right],
\end{equation}
where the expectation is taken over the point distribution $p \left(A | B, C \right)$ and $p \left(A \big| C \right)$.
With these definitions, the TE is defined as \cite{schreiber2000measuring}
\begin{align} \label{TE}
    \Phi_{\rm TE} & \left(\mathbf{Y}_i^D \rightarrow \mathbf{Y}_j^A \right) \notag\\
    & = I \left( Y_j^A[k];\mathbf{Y}_i^D[k-1:k-s] \big| \mathbf{Y}_j^A[k-1:k-r] \right),
\end{align}
where $s$ and $r$ are fixed integers; $\mathbf{Y}_j^A[k-1:k-r] = \left\{ Y_j^A[k-1], Y_j^A[k-2], ..., Y_j^A[k-r]\right\}$ and $\mathbf{Y}_i^D[k-1:k-s] = \left\{ Y_i^D[k-1], Y_i^D[k-2], ..., Y_i^D[k-s]\right\}$ denote windows of past samples for $\mathbf{Y}_j^A$ and $\mathbf{Y}_i^D$ respectively.
In practice, the TE is estimated using available data sequences $\mathbf{Y}_j^A$ and $\mathbf{Y}_i^D$. The TE was used for topology estimation in \cite{laghate2017learning, sharma2019communication}

\subsection{Setting the Threshold}

The threshold $\theta_{i,j}$ in (\ref{inference form}) can be set via a permutation test \cite{seth2010matlab}.
Accordingly, one considers a statistical significance test in which the null hypothesis corresponds to the assumption that the two sequences $\mathbf{Y}_i^D$ and $\mathbf{Y}_j^A$ are not causally related.
To obtain the distribution of the causality metrics $\Phi \left(\mathbf{Y}_i^D \rightarrow \mathbf{Y}_j^A \right)$ under the null hypothesis, $S$ random permutations of the sequences are obtained by considering permutations of the observed sequences.
$\mathbf{Y}_{i,s}^D$ and $\mathbf{Y}_{j,s}^A$ of sequences $\mathbf{Y}_i^D$ and $\mathbf{Y}_j^A$ are produced, with $s \in \{1,2,...,S\}$.
The causality metrics $\Phi \left( \mathbf{Y}_{i,s}^D \rightarrow \mathbf{Y}_{j,s}^A \right)$, with $s \in \{1,2,...,S\}$, are evaluated; and the threshold $\theta_{i,j}$ is set as the ($1 - \alpha$)-quantile of the empirical distribution of the samples $\left\{ \Phi \left( \mathbf{Y}_{i,s}^D \rightarrow \mathbf{Y}_{j,s}^A \right) \right\}_{s=1}^S$, when $\alpha \in [0, 1]$ is a fixed false alarm probability.

\section{EM-based Topology Estimation} \label{EM formulation}

The CDA schemes reviewed in the previous section were devised under the assumption that there are no packet losses \cite{tilghman2013inferring, laghate2017learning, sharma2019communication}.
As we argue in Sec. \ref{argue}, packet losses tend to make the CDA test (\ref{inference form}) unreliable, since the causality metrics are decreased in the presence of packet losses due to the missed association between data and ACK sequences erased by lost data packets.
To address this challenge, in this section, we introduce the EM-based CDA, which models packet losses using latent random variables.

\subsection{Impact of Packet Losses on CDA} \label{argue}

\begin{figure}[htp]
    \centering
    \includegraphics[scale=0.32]{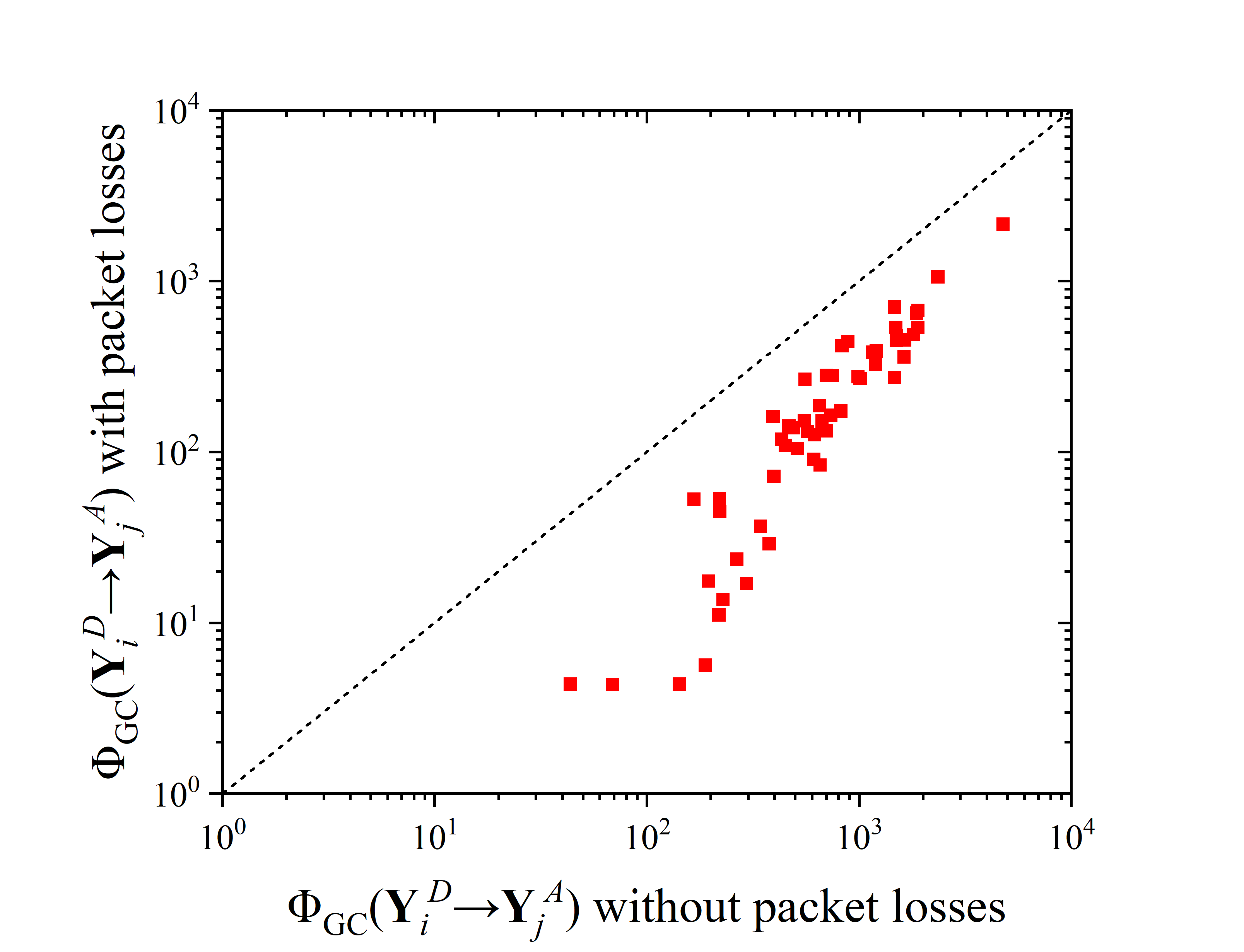}
    \caption{Causality metric (\ref{GCT}) evaluated for different links of an IEEE 802.11 ad-hoc network simulated on NS-3, with $N=12$ nodes, $M=65$ links, observation duration $60 \ {\rm s}$, time slot duration $T_s = 1.5 \ {\rm ms}$, and probability of packet loss 0.25.}
    \label{fig-causality loss}
\end{figure}

In order to gain insights into the impact of packet losses on the performance of CDA, we now consider an IEEE 802.11 ad-hoc network simulated with NS-3, and evaluate the GC metric (\ref{GCT}) for a given link $(i, j)$ in the presence and absence of packet losses. Details of the experimental setting can be found in Sec. \ref{NS-3}. Fig. \ref{fig-causality loss} reports the GC metric evaluated with losses as a function of the corresponding metric evaluated in a lossless scenario under the same conditions. Different points correspond to distinct links in the set $\mathcal{L}$.
The figure confirms that the GC metric tends to be decreased by packet losses, making CDA methods potentially ineffective.

\subsection{Parametric Model with Latent Variables} \label{PM}

The EM-Based Causality Discovery Algorithm (EM-CDA) scheme is based on the idea of formulating the problem of topology inference as the maximum likelihood estimate (MLE) of the adjacency matrix $\mathbf{A}$ in the presence of latent variables describing packet losses.
To elaborate, let $\mathbf{Y} = \left\{ \mathbf{Y}_{i}^{D}, \mathbf{Y}_{i}^{A} \big| \forall i \in \mathcal{N} \right\}$ be the observations.
We also introduce two sets of latent variables. The first, $\mathbf{D} = \left\{ D_{i,j}[k] \big| \forall i,j \in \mathcal{N},k \in \mathcal{K} \right\}$, contains variables $D_{i, j}[k]$ for all pairs of nodes $i$ and $j$ and time slots $k$, such that
\begin{equation}
    D_{i, j}[k] = 
    \left\{ \begin{array}{ll}
    1 & \mbox{if a data packet is sent on link $(i,j)$ in time slot $k$}, \\
    0 & \mbox{otherwise}.
    \end{array}\right.
\end{equation}
The second, $\mathbf{E} = \left\{ E_{i,j}[k] \big| \forall i,j \in \mathcal{N},k \in \mathcal{K} \right\}$, contains the packet loss variables defined in (\ref{error}). Note that the true value of the latent variables $D_{i,j}[k]$ and $E_{i,j}[k]$ are undefined for links not in set $\mathcal{L}$.
The set $\mathbf{Z} = \left\{ \mathbf{E}, \mathbf{D} \right\}$ defines the latent variables. Overall, we have observations $\mathbf{Y}$ and latent variables $\mathbf{Z}$.

We now define a parametric model that specifies the point distribution $p \left( \mathbf{Y}, \mathbf{Z} \big| \mathbf{\Theta} \right)$ of observations $\mathbf{Y}$ and latent variables $\mathbf{Z}$ as a function of a set of parameters, $\mathbf{\Theta}$.
Set $\mathbf{\Theta}$ includes the adjacency matrix $\mathbf{A}$, which is the quantity of interest, as well as some nuisance parameters to be introduced next.
We emphasize that the probabilistic model $p \left( \mathbf{Y}, \mathbf{Z} \big| \mathbf{\Theta} \right)$ does not generally describe the ground-truth data generation mechanism, which is unknown. Rather, it amounts to a set of assumptions made in order to develop the proposed topology estimation algorithm.

\begin{figure}[htp]
    \centering
    \includegraphics[scale=0.4]{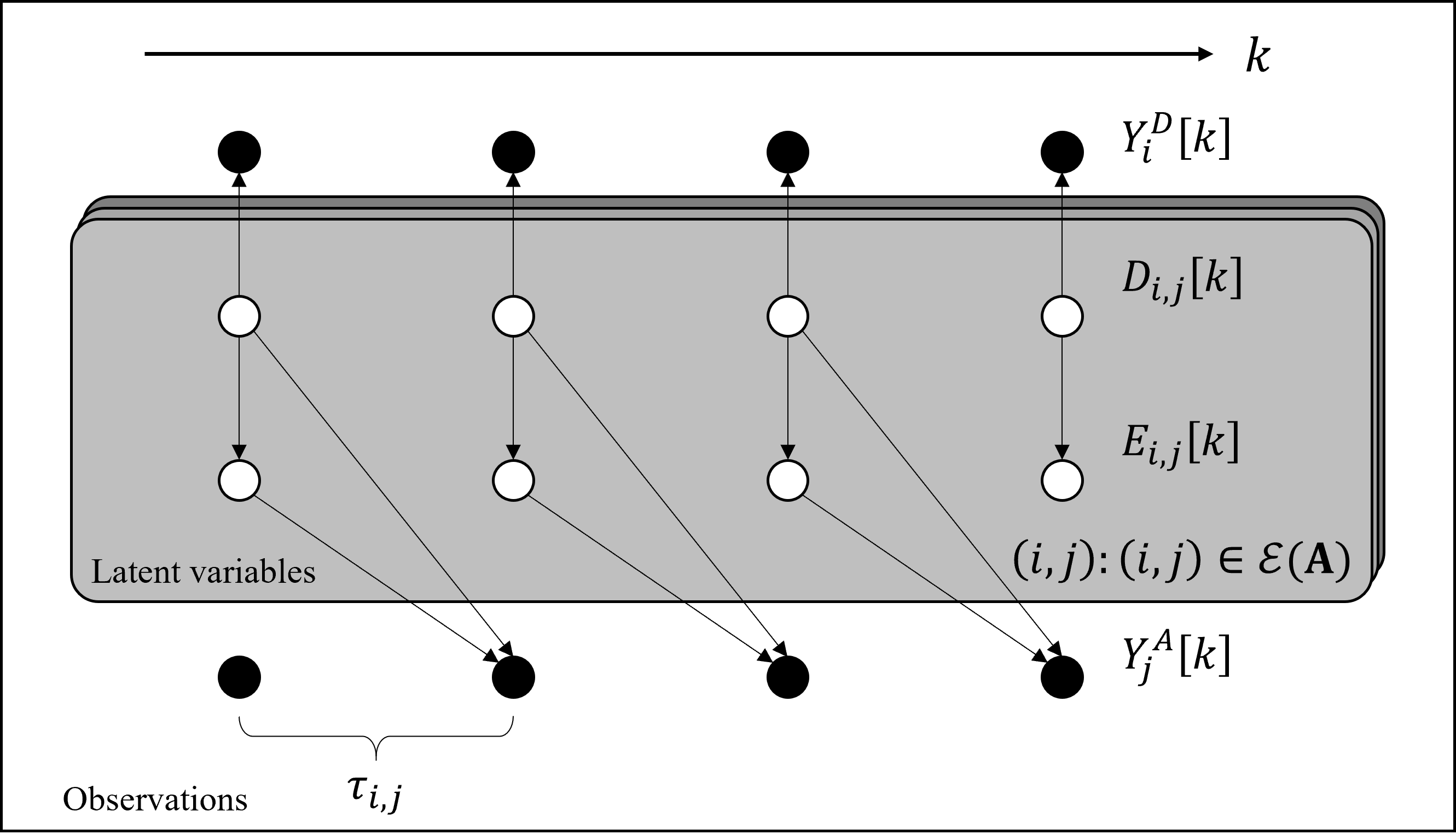}
    \caption{Bayesian network of the parametric model assumed in the derivation of EM-CDA. Shaded circles correspond to observed variables, and we set $\tau_{i,j}=1$ for simplicity of illustration. Note that the observations $Y_i^D[k]$ and $Y_j^A[k]$ depend only on latent variables indexed by $i$ and $j$, respectively, with $(i, j) \in \mathcal{E}(\mathbf{A})$.}
    \label{fig-Bayesian network}
\end{figure}

The parametric model, $p \left( \mathbf{Y}, \mathbf{Z} \big| \mathbf{\Theta} \right) = p ( \mathbf{Z} \big| \mathbf{\Theta} ) p ( \mathbf{Y} \big| \mathbf{Z}; \mathbf{\Theta} )$, depends on the set of unknown parameters $\mathbf{\Theta} = \left\{\mathbf{A}, \mathbf{L}, \mathbf{R}, \mathbf{T} \right\}$, where matrices $\left\{ \mathbf{L}, \mathbf{R}, \mathbf{T} \right\}$ are nuisance parameter matrices representing error rate, transmission rate, and ACK delay on each link, respectively. Let us define as $\mathcal{E}(\mathbf{A})$ the set of coordinates of non-zero entries of the adjacency matrix $\mathbf{A}$, that is, the estimated links given matrix $\mathbf{A}$. To start, we assume that variables $(E_{i, j}[k], D_{i, j}[k])$ corresponding to different link $(i, j)$ are independent, i.e.,
\begin{align} \label{em-0}
    p \left( \mathbf{Z} \big| \mathbf{\Theta} \right)
    = \prod_{(i, j) \in \mathcal{E}(\mathbf{A})} p\left( \mathbf{E}_{i, j}, \mathbf{D}_{i, j} \big| \mathbf{\Theta} \right),
\end{align}
where we have the sequences $\mathbf{E}_{i, j} = {\left\{ E_{i, j}[k] \right\}}_{k=1}^K$ and $\mathbf{D}_{i, j} = {\left\{ D_{i, j}[k] \right\}}_{k=1}^K$.
Focusing now on sequences $\mathbf{E}_{i, j}$ and $\mathbf{D}_{i, j}$, we assume the joint distribution
\begin{align} \label{em-1}
    p\left( \mathbf{E}_{i, j}, \mathbf{D}_{i, j} \big| \mathbf{\Theta} \right)
    & = \prod_{k \in \mathcal{K}} \Big[ p \left( D_{i, j}[k] \big| \mathbf{D}_{i, j}[1:k-1], \mathbf{E}_{i, j}[1:k-1]; \mathbf{\Theta} \right) \notag\\
    & \quad \quad \times p\left( E_{i, j}[k] \big| \mathbf{D}_{i, j}[1:k], \mathbf{E}_{i, j}[1:k-1]; \mathbf{\Theta} \right) \Big] \notag\\
    & = \prod_{k \in \mathcal{K}} \Big[ p\left( D_{i, j}[k] \big| R_{i, j} \right) p\left( E_{i, j}[k] \big| D_{i, j}[k]; L_{i, j} \right) \Big],
\end{align}
where the first equality follows from the chain rule of probability, and the second is a consequence of the following two assumptions. First, we assume that an error on a link $(i, j)$, indicated by $E_{i, j}[k]=1$, occurs with probability $L_{i, j}$ if a transmission occurred on the same link, i.e., if $D_{i, j}[k]=1$. This is expressed with the conditional distribution
\begin{align} \label{p-e}
    p &\left( E_{i, j}[k] \big| D_{i, j}[k]; L_{i, j} \right) \notag\\
    & =
    \left\{ \begin{array}{ll}
    L_{i, j}^{E_{i, j}[k]}(1 - L_{i, j})^{1 - E_{i, j}[k]} & \mbox{if $D_{i, j}[k] = 1$}, \\
    1 - E_{i, j}[k] & \mbox{otherwise}.
    \end{array}\right.
\end{align}
Second, transmissions occur independently of previous transmissions and errors with probability $R_{i, j}$, which is formulated as
\begin{align} \label{p-d}
    p \left( D_{i, j}[k] \big| R_{i, j} \right)
    = R_{i, j}^{D_{i, j}[k]}(1 - R_{i, j})^{1 - D_{i, j}[k]}.
\end{align}
We emphasize that the conditional distribution (\ref{p-d}) entails a significant approximation, since transmissions in many network scenarios encompass also retransmission of previous, erroneously received, packets. The Bayesian network that describes the assumed model for the latent variables is shown in Fig. \ref{fig-Bayesian network}.

To fully specify the parametric model $p \left( \mathbf{Y}, \mathbf{Z} \big| \mathbf{\Theta} \right) = p ( \mathbf{Z} \big| \mathbf{\Theta} ) p ( \mathbf{Y} \big| \mathbf{Z}; \mathbf{\Theta} )$ we need to describe also the distribution $p \left( \mathbf{Y} \big| \mathbf{Z}; \mathbf{\Theta} \right)$.
In this regard, the observations $\mathbf{Y}$ are assumed to be a function $f\left(\mathbf{D}, \mathbf{E} \big| \mathbf{A}, \mathbf{T}\right)$ of the latent variables $\mathbf{D}$ and $\mathbf{E}$ that is parameterized by the adjacency matrix $\mathbf{A}$ and the matrix of delays $\mathbf{T}$. Accordingly, the distribution of $\mathbf{Y}$ conditioned on $\mathbf{Z}$ is given by
\begin{equation} \label{post-dis}
    p\left( \mathbf{Y} \big| \mathbf{Z}; \mathbf{\Theta} \right) = p\left( \mathbf{Y} \big| \mathbf{D}, \mathbf{E}; \mathbf{A}, \mathbf{T} \right) = \delta(\mathbf{Y} - f\left(\mathbf{D}, \mathbf{E} \big| \mathbf{A}, \mathbf{T}\right)),
\end{equation}
where $\delta(\cdot)$ is the Kronecker delta function. Function $f\left(\mathbf{D}, \mathbf{E} \big| \mathbf{A}, \mathbf{T}\right)$ is defined as follows.
Since the number of packets observed from a node $i$ equals the sum of the numbers of packets sent to other nodes $j$ with $(i, j) \in \mathcal{E}(\mathbf{A})$ in the given time slot, we have the equality
\begin{equation} \label{model-data}
    Y_i^D[k] = \sum_{j: (i, j) \in \mathcal{E}(\mathbf{A})}D_{i, j}[k].
\end{equation}
This is reflected by the Bayesian network in Fig. \ref{fig-Bayesian network}.
Similarly, the number of ACKs reported by a node $j$ is equal to the sum of the numbers of ACKs sent to other nodes $i$ with $(i, j) \in \mathcal{E}(\mathbf{A})$. Defining the model parameter $\tau_{i, j}$ as the delay between ACK and packet transmission on link $(i, j)$, we thus assume the equality
\begin{equation} \label{model-arq}
    Y_j^A[k] = \sum_{i: (i, j) \in \mathcal{E}(\mathbf{A})} (1 - E_{i, j}[k - \tau_{i, j}])D_{i, j}[k - \tau_{i, j}].
\end{equation}
It is recalled that, while the actual unknown time delays $\tau_{i, j}[k]$ in (\ref{DATA}) may depend on the time slot $k$, the parameters $\tau_{i, j}$ in (\ref{model-arq}), which are collected in matrix $\mathbf{T}$, are assumed to be static in order to facilitate estimation.
Overall, equalities (\ref{model-data})-(\ref{model-arq}) define function $f\left(\mathbf{D}, \mathbf{E} \big| \mathbf{A}, \mathbf{T}\right)$ and hence distribution (\ref{post-dis}).


\subsection{EM-Based Causality Discovery Algorithm (EM-CDA)} \label{EM}

Given the likelihood $p \left( \mathbf{Y}, \mathbf{Z} \big| \mathbf{\Theta} \right)$ of the complete data $(\mathbf{Y}, \mathbf{Z})$, EM-CDA aims to address the MLE problem
\begin{equation} \label{ML}
    \mathop{\mathrm{max}}\limits_{\mathbf{\Theta}}\left\{p \left( \mathbf{Y} \big| \mathbf{\Theta} \right)
    = \mathbb{E}_{p (\mathbf{Z} | \mathbf{\Theta})} \left[ p \left(\mathbf{Y} \big| \mathbf{Z}; \mathbf{\Theta} \right) \right] \right\}
\end{equation}
via EM.
Accordingly, EM-CDA updates the current estimate $\mathbf{\Theta}$ across a number of iteration, producing a sequence of iterates $\mathbf{\Theta}^{(1)}, \mathbf{\Theta}^{(2)}, ..., \mathbf{\Theta}^{(n)}$.
At each iteration $n$, EM first performs the expectation step (E-step), which evaluates the expected value
\begin{align} \label{origin-Q}
    Q \left(\mathbf{\Theta} \big| \mathbf{\Theta}^{(n)} \right)
    = \mathbb{E}_{p ( \mathbf{Z} | \mathbf{Y}; \mathbf{\Theta}^{(n)} )} \left[ \log p\left( \mathbf{Y}, \mathbf{Z} \big| \mathbf{\Theta} \right) \right]
\end{align}
of the complete log-likelihood $\log p\left( \mathbf{Y}, \mathbf{Z} \big| \mathbf{\Theta} \right)$ with respect to the current posterior distribution $p \left( \mathbf{Z} | \mathbf{Y}; \mathbf{\Theta}^{(n)} \right)$.
Then, the maximization step (M-step) is carried out, wherein the next update is obtained as
\begin{equation} \label{argmax}
    \mathbf{\Theta}^{(n+1)} = \mathop{\mathrm{argmax}}\limits_{\mathbf{\Theta}}{Q \left(\mathbf{\Theta} \big| \mathbf{\Theta}^{(n)} \right)}.
\end{equation}

A direct application of EM to the model $p \left( \mathbf{Y}, \mathbf{Z} \big| \mathbf{\Theta} \right)$ described in the previous subsection is computationally infeasible. To obtain a scalable solution, EM-CDA approximates the E-step using Monte Carlo sampling, and the M-step via CDA (see Sec. \ref{Causality method}).
The resulting algorithm can be viewed as an iterative generalization of CDA, wherein estimates of packet losses are accounted for in the estimates of the causality metrics in order to address the issue described in Sec. \ref{argue}.
We detail both E-step and M-step in the rest of this section, and the overall EM-CDA is described in Algorithm \ref{alg:Framework}.

\subsection{Expectation Step (E-step)} \label{E-step}

At iteration $n$, given the current parameters $\mathbf{\Theta}^{(n)}$, the E-step aims at generating $M$ samples $\{\mathbf{Z}_1^{(n)}, ..., \mathbf{Z}_M^{(n)}\}$ from the posterior distribution $p \left( \mathbf{Z} \big| \mathbf{Y}; \mathbf{\Theta}^{(n)} \right)$. With such samples, the function $Q \left(\mathbf{\Theta} \big| \mathbf{\Theta}^{(n)} \right)$ in (\ref{origin-Q}) is approximated via the stochastic estimate \cite{liu2019parameter}
\begin{align} \label{SAEM}
    Q \left(\mathbf{\Theta} \big| \mathbf{\Theta}^{(n)} \right)
    & = \left(1 - \gamma^{(n)} \right) Q \left(\mathbf{\Theta} \big| \mathbf{\Theta}^{(n-1)} \right) \notag\\
    & \quad + \frac{\gamma^{(n)}}{M} \sum_{m=1}^M \log p\left( \mathbf{Y}, \mathbf{Z}_m^{(n)} \big| \mathbf{\Theta} \right),
\end{align}
where $\gamma^{(n)} \in [0, 1]$ is a learning rate.

In order to generate the samples $\mathbf{Z}_m^{(n)} \sim p \left( \mathbf{Z} \big| \mathbf{Y}; \mathbf{\Theta}^{(n)} \right)$ for $m = 1, ..., M$, we apply Gibbs sampling. Gibbs sampling generates the samples $\mathbf{Z}_m^{(n)}$ sequentially over index $m=1, ..., M$ by drawing samples from the conditional probabilities of one variable in $\mathbf{Z}$ given all other variables in $\mathbf{Z}$ \cite{koller2009probabilistic}.
Accordingly, each sample $\mathbf{Z}_{m}^{(n)} = \left\{ D_{i, j, m}^{(n)}[k], E_{i, j, m}^{(n)}[k] \big| \forall i, j \in \mathcal{N}, k \in \mathcal{K} \right\}$ is generated as follows.

Using the notations $Z_{i,j}[k] = (D_{i,j}[k],E_{i,j}[k])$ and $Z_{-(i,j)}[-k] = \{D_{i,j}[k], E_{i,j}[k]\}_{\substack{(i^\prime, j^\prime) \neq (i,j) \\ k^\prime \neq k}}$, for each pair of variables $Z_{i,j}[k]$, we sample from the posterior $p \left( Z_{i,j}[k] \big| Z_{-(i,j)}[-k], \mathbf{Y}; \mathbf{\Theta}^{(n)} \right)$ given all other variables. This can be evaluated as
\begin{align} \label{sample-1}
    p & \left(Z_{i,j}[k] \big| Z_{-(i,j)}[-k], \mathbf{Y}; \mathbf{\Theta}^{(n)} \right) \notag\\
    & = p\left(Z_{i,j}[k] \big| Y_{i}^D[k], Y_j^A[k+\tau_{i, j}]; \mathbf{\Theta}^{(n)} \right) \notag\\
    & = \frac{p\left(Z_{i,j}[k], Y_{i}^D[k], Y_j^A[k+\tau_{i, j}] \big| \mathbf{\Theta}^{(n)}\right)}
    {p\left(Y_{i}^D[k], Y_j^A[k+\tau_{i, j}] \big| \mathbf{\Theta}^{(n)}\right)} \notag\\
    & = \frac{p\left(Z_{i,j}[k] \big| \mathbf{\Theta}^{(n)}\right) p\left(Y_{i}^D[k], Y_j^A[k+\tau_{i, j}] \big| Z_{i, j}[k]; \mathbf{\Theta}^{(n)}\right)}
    {\sum_{Z_{i, j}[k]} p\left(Z_{i,j}[k] \big| \mathbf{\Theta}^{(n)} \right) p\left(Y_{i}^D[k], Y_j^A[k+\tau_{i, j}] \big| Z_{i, j}[k]; \mathbf{\Theta}^{(n)}\right)},
\end{align}
where the first equality follows from d-separation based on the Bayesian network in Fig. \ref{fig-Bayesian network} (see, e.g., \cite{simeone2018brief}), and $p\left(Z_{i,j}[k] \big| \mathbf{\Theta}^{(n)}\right)$ is given by the product of (\ref{p-e}) and (\ref{p-d}) as
\begin{align} \label{sample-2}
    p\left(Z_{i,j}[k] \big| \mathbf{\Theta}^{(n)}\right)
    = p\left( D_{i, j}[k] \big| R_{i, j}^{(n)} \right) p\left( E_{i, j}[k] \big| D_{i, j}[k]; L_{i, j}^{(n)} \right).
\end{align}
We now left with the problem evaluating the distribution $p\left(Y_{i}^D[k], Y_j^A[k+\tau_{i, j}] \big| Z_{i, j}[k]; \mathbf{\Theta}^{(n)}\right)$. According to (\ref{model-data})-(\ref{model-arq}), it is given by the probability of that $Y_{i}^D[k] - D_{i, j}[k]$ packets are sent by node $i$ to other nodes except $j$ at time $k$, and that $Y_j^A[k+\tau_{i, j}] - D_{i, j}[k](1 - E_{i, j}[k])$ ACKs are sent by node $j$ to other nodes except $i$ at time $k+\tau_{i, j}$.
Therefore, by (\ref{em-1})-(\ref{p-d}) we have
\begin{align} \label{sample-3}
    p & \left(Y_{i}^D[k], Y_j^A[k+\tau_{i, j}] \big| Z_{i, j}[k]; \mathbf{\Theta}^{(n)}\right) \notag\\
    & = p \left(Y_{i}^D[k] \big| Z_{i, j}[k]; \mathbf{\Theta}^{(n)}\right) p \left(Y_j^A[k+\tau_{i, j}] \big| Z_{i, j}[k]; \mathbf{\Theta}^{(n)}\right) \notag\\
    & =  {\rm Bin} \left( Y_{i}^D[k] - D_{i, j}[k] \big| {\{R_{i,j}^{(n)}\}}_{\substack{(i, l) \in \mathcal{E}(\mathbf{A}) \\ l \neq j}} \right) \notag\\
    & \quad \times {\rm Bin} \left( Y_j^A[k+\tau_{i, j}] - D_{i, j}[k](1 - E_{i, j}[k]) \big| {\{R_{i,j}^{(n)} (1 - L_{i,j}^{(n)})\}}_{\substack{(l, j) \in \mathcal{E}(\mathbf{A}) \\ l \neq i}} \right),
\end{align}
where we denote as ${\rm Bin}(y | {\{p_i\}}_{i=1}^L)$ the probability mass function of a sum of $L$ independent Bernoulli random variables, with each $i$th random variables having probability $p_i$ of being equal to 1.

\subsection{Maximization Step (M-step)}

Given $\mathbf{Y}$ and samples generated $\{\mathbf{Z}_1^{(n)}, ..., \mathbf{Z}_M^{(n)}\}$ in the E-step, the M-step aims at updating parameters $\mathbf{\Theta}$.
The discrete parameters $\mathbf{T}$ and $\mathbf{A}$ are updated by generalizing the CDA approach described in Sec. \ref{Causality method} to include the estimate of delays.
The continuous parameters $\mathbf{R}$ and $\mathbf{L}$ are then updated by finding the stationary points of the objective function of $Q \left(\mathbf{\Theta} \big| \mathbf{\Theta}^{(n)} \right)$ in (\ref{SAEM}).

For each sample $\mathbf{Z}_m^{(n)}$, we define as $\mathbf{Y}_{i,m}^{D, (n)} = \left\{\sum_{j \in \mathcal{N}}D_{i, j, m}^{(n)}[k]  \big| \forall k \in \mathcal{K} \right\}$ the estimated data packet sequence for node $i$;
and as $\mathbf{Y}_{j, m}^{A, (n)} = \left\{ Y_j^A[k] + \sum_{(i, j) \in \mathcal{E}(\mathbf{A})} E_{i, j, m}^{(n)}[k-\tau_{i, j}^{(n)}] \big| \forall k \in \mathcal{K} \right\}$ the estimated ACK sequence for node $j$.
To update the delay matrix for $\mathbf{T}_m^{(n)} = \left\{ \tau_{i,j,m}^{(n)} \big| \forall i,j \in \mathcal{N} \right\}$, we obtain the sequences $\mathbf{Y}_{i, m}^{D, (n), \tau} = {\{ Y_{i, m}^{D, (n)}[k + \tau] \}}_{k=1}^K$ by shifting backward in time by $\tau$ steps the sequences $\mathbf{Y}_{i, m}^{D, (n)}$. Then, the causal dependence measure $\Phi (\mathbf{Y}_{i, m}^{D, (n), \tau} \rightarrow \mathbf{Y}_{j, m}^{A, (n)} )$ is calculated using (\ref{GCT}) or (\ref{TE}) for a range of values $[1, \tau_{max}]$ to obtain the estimate
\begin{equation} \label{up-delay}
    \tau_{i, j, m}^{(n)} = \mathop{\mathrm{argmax}}\limits_{\tau \in [1, \tau_{max}]}{\Phi \left(\mathbf{Y}_{i, m}^{D, (n), \tau} \rightarrow \mathbf{Y}_{j, m}^{A, (n)} \right)}.
\end{equation}

Furthermore, using (\ref{inference form}), the estimated topology entries $a_{i,j,m}^{(n)}$ of the adjacency matrix $\mathbf{A}_m^{(n)}$ are given by
\begin{equation} \label{up-a}
    a_{i,j,m}^{(n)} = 
    \left\{ \begin{array}{ll}
    1 & \mbox{if $\Phi  ( \mathbf{Y}_{i,m}^{D, (n)} \rightarrow \mathbf{Y}_{i, m}^{A, (n)} ) > \theta_{i, j, m}^{(n)}$}, \\
    0 & \mbox{otherwise},
    \end{array}\right.
\end{equation}
where $\theta_{i, j, m}^{(n)}$ is a threshold. Then, we set
\begin{equation} \label{up-A}
    a_{i,j}^{(n+1)} = 
    \left\{ \begin{array}{ll}
    1 & \mbox{if $\sum_{m=1}^M a_{i,j,m}^{(n)} \geq \frac{M}{2}$}, \\
    0 & \mbox{otherwise},
    \end{array}\right.
\end{equation}
that is, an edge $(i, j)$ is included in the set $\mathcal{E}(\mathbf{A}^{(n)})$ of the majority of tests (\ref{up-a}) set $a_{i,j,m}^{(n)} = 1$.

Finally, setting the partial derivatives of $Q \left(\mathbf{\Theta} \big| \mathbf{\Theta}^{(n)} \right)$ in (\ref{SAEM}) with respect to $\mathbf{R}$ and $\mathbf{L}$ to zero, respectively, the updated $\mathbf{R}^{(n+1)}$ and $\mathbf{L}^{(n+1)}$ are given by the empirical averages as
\begin{align} \label{up-R}
    R_{i,j}^{(n+1)}
    = \left(1 - \gamma^{(n)} \right) R_{i,j}^{(n)}
    + \frac{\gamma^{(n)}}{MK} \sum_{m=1}^M \sum_{k \in \mathcal{K}} D_{i,j,m}^{(n)}[k],
    \quad \forall (i,j) \in \mathcal{E}(\mathbf{A}^{(n)}),
\end{align}
and
\begin{align} \label{up-L}
    L_{i,j}^{(n+1)}
    = \left(1 - \gamma^{(n)} \right) L_{i,j}^{(n)}
    + \gamma^{(n)} \frac{\sum_{m=1}^M \sum_{k \in \mathcal{K}} E_{i,j,m}^{(n)}[k]}{\sum_{m=1}^M \sum_{k \in \mathcal{K}} D_{i,j,m}^{(n)}[k]}, \quad \forall (i,j) \in \mathcal{E}(\mathbf{A}^{(n)}).
\end{align}

\begin{algorithm}[htp] 
  \caption{EM-CDA}
  \label{alg:Framework}  
  \begin{algorithmic}[1]  
    \Require 
      The observations $\mathbf{Y}$, learning rate sequences $\{ \gamma^{(n)} \}$, number of samples $M$, maximum estimated delay $\tau_{max}$, and significance level $\alpha$;
    \Ensure 
      Matrix of inferred communication links $\mathbf{\hat{A}}$;
      
    \State \textbf{Initialization:} Initialize  $\mathbf{L}^{(0)}$ and $\mathbf{R}^{(0)}$ with 0-1 uniform distribution; the adjacency matrix $\mathbf{A}^{(0)}$ to have every entry equal to one; $\mathbf{T}^{(0)}$ by (\ref{up-delay}) using $\mathbf{Y}$; $n$ to 0;
    \label{code:fram:init1}
    
    \State \textbf{while} $\mathbf{\Theta}^{(n)}$ has not converged \textbf{do}
    \State \quad Generate samples based on (\ref{sample-1})-(\ref{sample-3});
    \State \quad Update $\mathbf{\Theta}^{(n+1)}$ based on (\ref{up-delay})-(\ref{up-L});
    \State \quad $n \leftarrow n+1$;
    \State \textbf{end while}
    \State Obtain $\mathbf{\hat{A}} = \mathbf{A}^{(n)}$. 
  \end{algorithmic}  
\end{algorithm}

\section{Numerical Results} \label{Numerical Results}
In this section, numerical results are provided to demonstrate the performance of the proposed EM-CDA scheme as corresponds to the conventional CDA methods reviewed in Sec. \ref{Causality method} \cite{tilghman2013inferring, laghate2017learning, sharma2019communication}.
We first consider a toy example in which we can evaluate the impact of the approximations adopted in the derivation of EM-CDA via an exact implementation of EM. Then, large-scale experiments are conducted by simulating wireless networks via NS-3 \cite{lacage2006yet, campanile2020computer}.

\subsection{Small-Scale Experiments}

In this subsection, we compare EM-CDA with an implementation of EM to address the MLE problem (\ref{ML}) that applies the exhaustive search (ES) method in the M-step to maximize the function $Q \left(\mathbf{\Theta} \big| \mathbf{\Theta}^{(n)} \right)$ over variables $\mathbf{A}$ and $\mathbf{T}$. We refer to this scheme as EM-ES.
To enable EM-ES over the exponential number of possible choices $\mathbf{A}$, we consider a small network with $N = 4$ nodes that is allowed to follow the same model adopted for the derivation of EM as explained in Sec. \ref{EM formulation}. In the next subsection, we will consider a more realistic scenario in NS-3.

Half of the links are randomly selected to be active; the ground-truth average transmission rate $R_{i,j}^{*}$ for all active links is set as 0.1; the average packet loss rate $L_{i,j}^{*}$ for all links is set to 0.05 or 0.5; and the ground-truth delay $\tau_{i,j}^{*}$ are set to 1 time slot. We simulate the network for 5000 time slots.
In the E-steps of both methods, the number of samples is set as $M=30$. In M-step, GCT or TE is adopted as the causality discovery algorithm in EM-CDA. The significance level $\alpha$ in (\ref{inference form}) is set to 0.05 as in \cite{sharma2019communication}. All the results are generated in 20 trials with different random initial values.

The probability of false alarm, $p_{\rm FA}$, and the probability of detection, $p_{\rm D}$, are adopted to measure the performance of topology inference. These metrics are defined as
\begin{equation}
    P_{\rm FA} = 
    \frac{{\rm FP}}{{\rm FP + TN}},
\end{equation}
and
\begin{equation}
    P_{\rm D} = \frac{{\rm TP}}{{\rm TP + FN}},
\end{equation}
where ${\rm TP}$ denotes the number of correctly detected existing links, ${\rm FN}$ denotes the number of missed existing links, ${\rm TN}$ denotes the number of correctly detected missing links, and ${\rm FP}$ denotes the number of incorrectly detected missing links.

\begin{figure}[htp]
    \centering
    \subfigure[]{
    \includegraphics[scale=0.3]{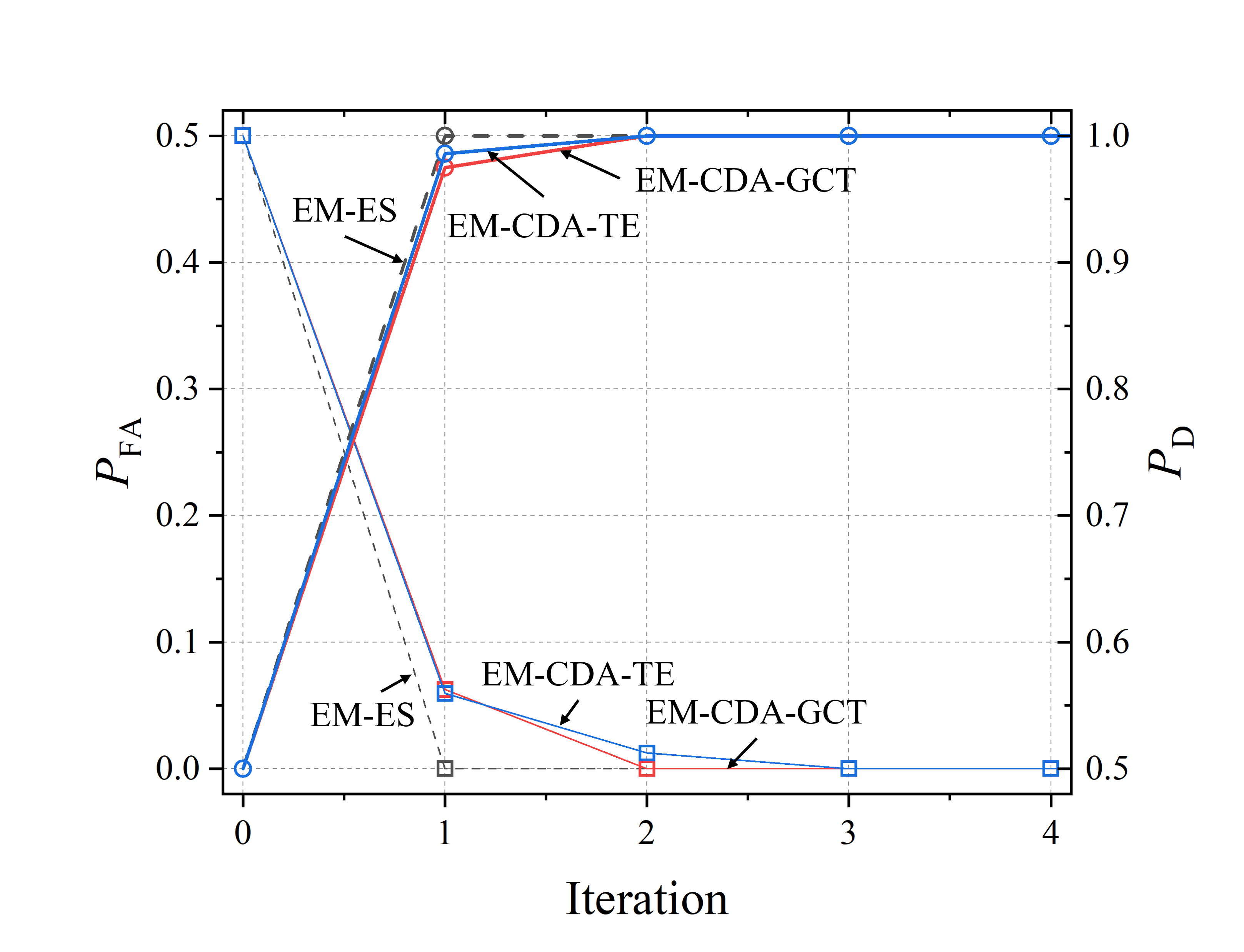}
    }
    \subfigure[]{
    \includegraphics[scale=0.3]{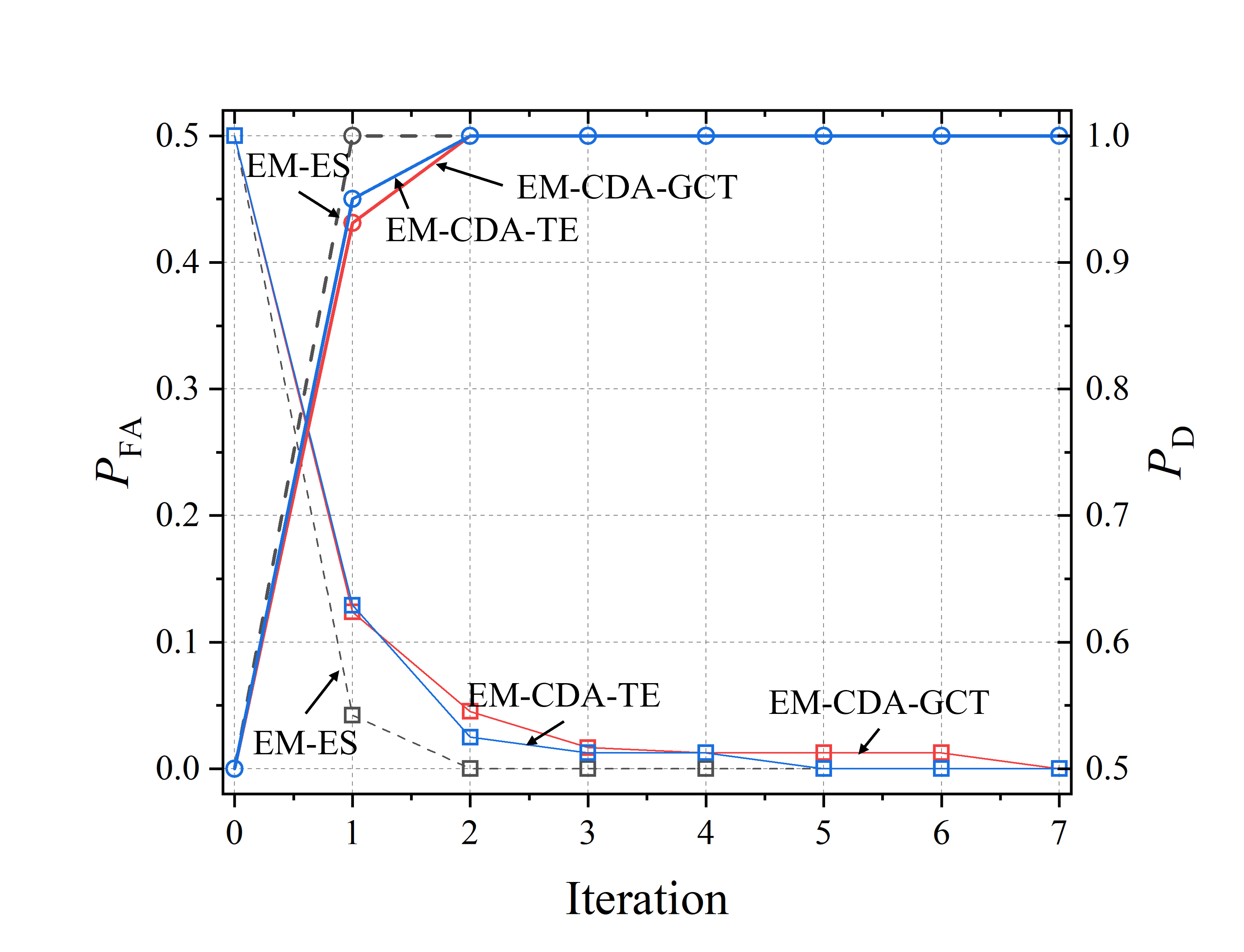}
    }
    \caption{Probability of false alarm $P_{\rm FA}$ and probability of detection $p_{\rm D}$ for topology inference versus the number of EM iterations for the ideal EM-ES scheme and for EM-CDA with GCT and TE causality metrics: iterations in the case of (a) $L_{i,j}^{*}=0.05$, and (b) $L_{i,j}^{*}=0.5$.}
    \label{fig-small}
\end{figure}
Fig. \ref{fig-small} shows the probabilities $P_{\rm FA}$ and $P_{\rm D}$ across the EM iterations.
EM-ES is seen to obtain the optimal solution, yielding the ideal case $P_{\rm FA} = 0$ and $P_{\rm D} = 1$, in a single iteration, while EM-CDA with both GCT and TE requires more iterations, but it is able to converge to the optimal solution.
Furthermore, the number of required EM iterations for the performance of EM-CDA increases as the ground-truth average loss rate $L_{i, j}^{*}$ increases, because, as discussed in Sec. \ref{argue}, unreliable observations provide missing and spurious information that needs to be compensated for by refining the estimates of the latent variables.

\subsection{Simulations on NS-3} \label{NS-3}

In this subsection, we test EM-CDA in different wireless scenarios simulated on NS-3 using the parameters in Table \ref{tab-ns3}.
\begin{table}[htp]
\caption{\centering Parameters Values for NS-3 Simulations}
\label{tab-ns3}
\centering
    \begin{tabular}{ccc}
    \hline\hline
    
    \specialrule{0em}{1pt}{1pt}
    \textbf{Parameter} & \textbf{Value} \\
    \specialrule{0em}{1pt}{1pt}
    
    \hline
    
    \specialrule{0em}{1pt}{1pt}
    Area size & 100 ${\rm m}^2$ \\
    \specialrule{0em}{1pt}{1pt}
    Carrier frequency $f_0$  & 2.412 GHz \\
    \specialrule{0em}{1pt}{1pt}
    Data packet size  & 1024 Bytes \\
    \specialrule{0em}{1pt}{1pt}
    MAC ACK size  & 36 Bytes \\
    \specialrule{0em}{1pt}{1pt}
    Channel packet loss rate & varies \\
    \specialrule{0em}{1pt}{1pt}
    Transmission rate & varies \\
    \specialrule{0em}{1pt}{1pt}
    Simulation duration $T$  & 60 s \\
    \specialrule{0em}{1pt}{1pt}
    Time slot duration $T_s$  & 1.5 ms \\
    \specialrule{0em}{1pt}{1pt}
    
    \hline\hline
    \end{tabular}
\end{table}
The system consists of $N$ nodes randomly and uniformly distributed within a 10 ${\rm m}$ $\times$ 10 ${\rm m}$ area that follow the an IEEE $802.11$ ad-hoc protocol operating at carrier frequency $f_0 = 2.412 \ {\rm GHz}$. Omnidirectional antennas are used at the nodes, with path-loss, log-normal shadowing, and thermal noise accounted for as in \cite{lacage2006yet}. The simulation lasts $T = 60 \ {\rm s}$, and the time slot duration is $T_s = 1.5 \ {\rm ms}$.
The offered traffic for each link is $1 \ {\rm Mbps}$, with a data packet size of 1024 Bytes and an ACK size of 36 Bytes. If not stated otherwise, we set $N = 12$ nodes, fraction of active links $0.5$, and average packet loss rate $0.3$.

\begin{figure}[htp]
    \centering
    \subfigure[]{
    \includegraphics[scale=0.3]{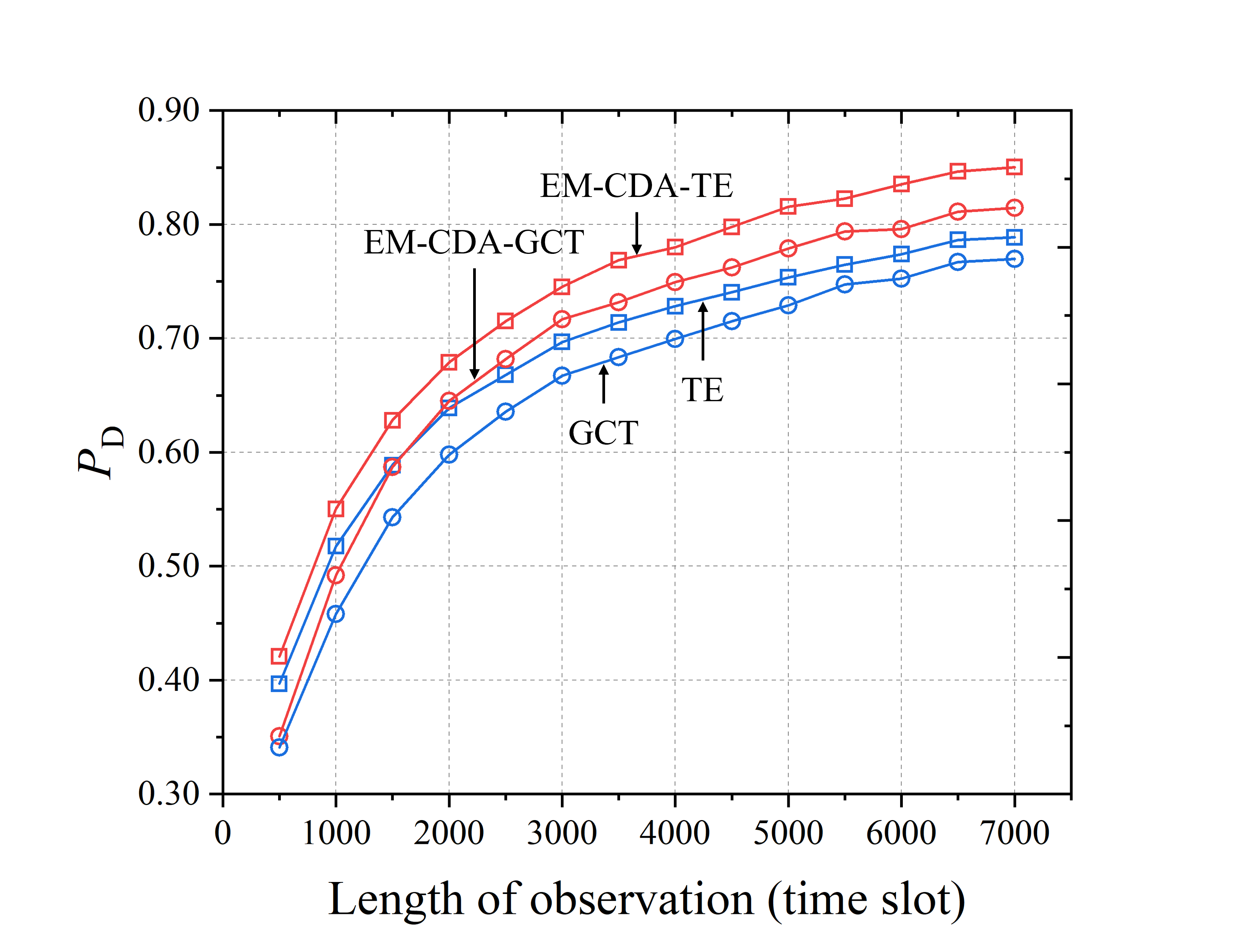}
    }
    \subfigure[]{
    \includegraphics[scale=0.3]{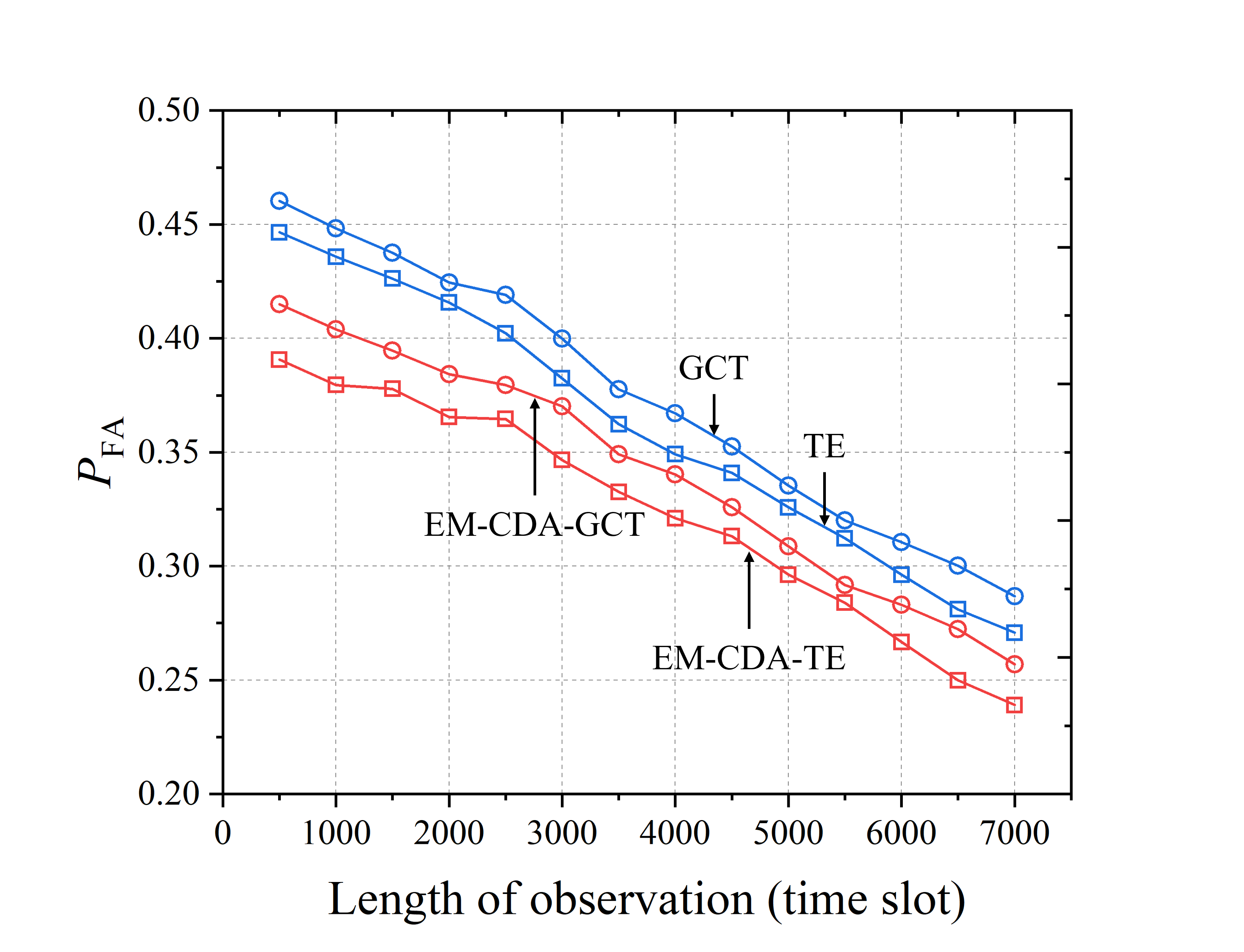}
    }
    \caption{(a) Probability of detection $p_{\rm D}$ and (b) probability of false alarm $P_{\rm FA}$ for topology inference versus the length of observation in time slots for CDA and EM-CDA with GCT and TE causality metrics.}
    \label{fig-length}
\end{figure}

Fig. \ref{fig-length} depicts the probabilities $P_{\rm FA}$ and $P_{\rm D}$ as a function of the number of observed time slots.
As more data are collected, EM-CDA is able to outperform CDA methods in terms of both probabilities, with gains saturating when enough information is collected.

\begin{figure}[htp]
    \centering
    \subfigure[]{
    \includegraphics[scale=0.3]{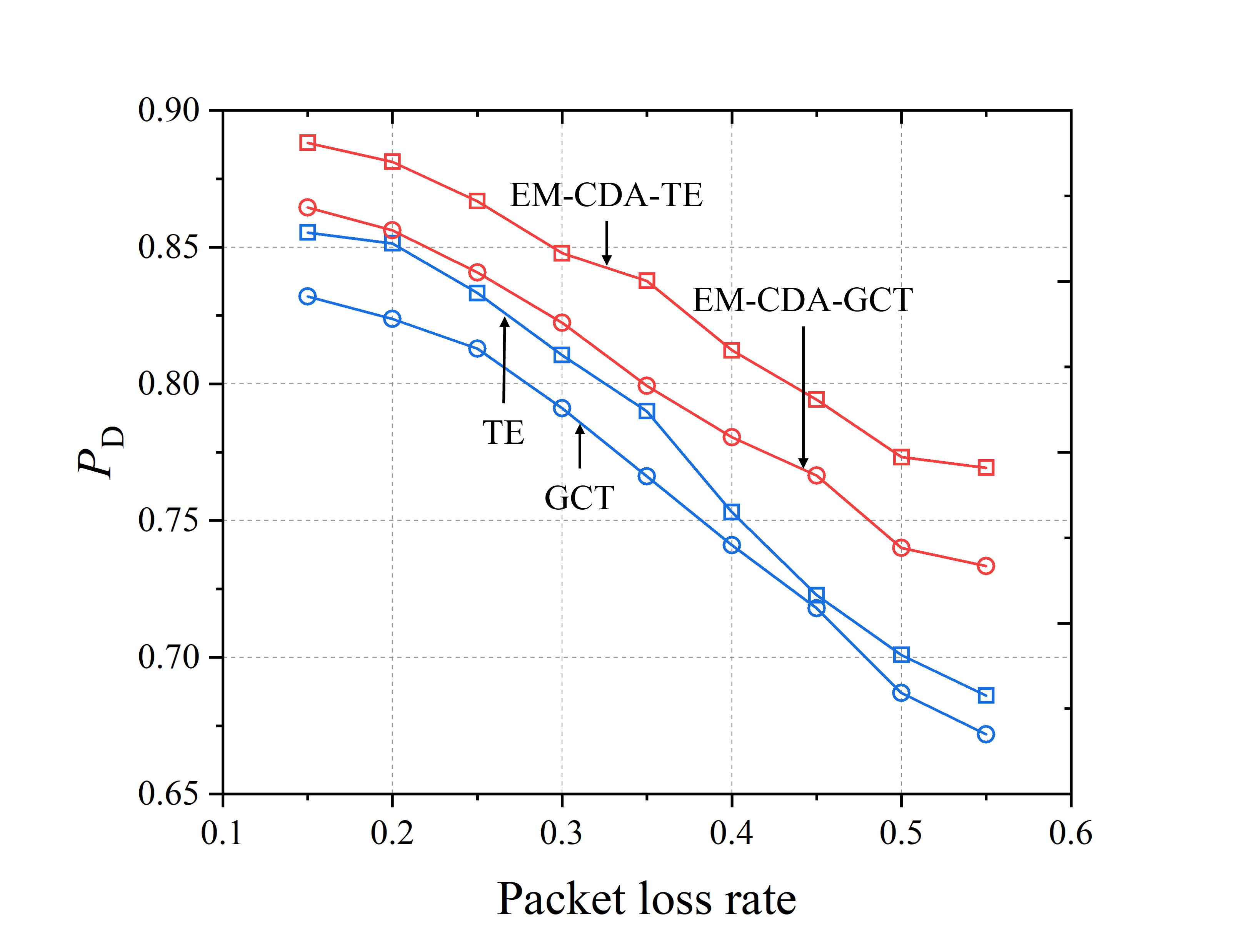}
    }
    \subfigure[]{
    \includegraphics[scale=0.3]{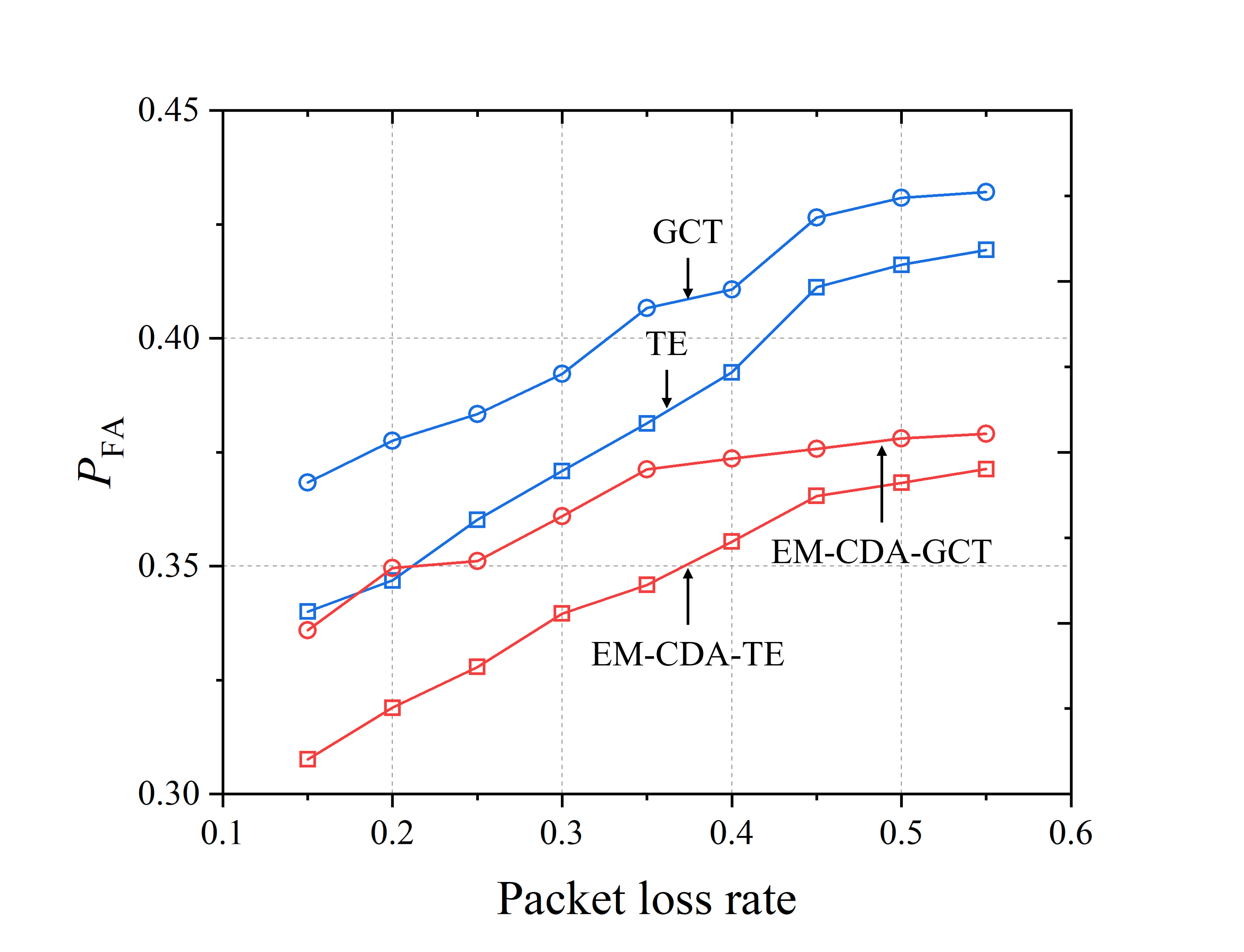}
    }
    \caption{(a) Probability of detection $p_{\rm D}$ and (b) probability of false alarm $P_{\rm FA}$ for topology inference versus the packet loss rate for CDA and EM-CDA with GCT and TE causality metrics.}
    \label{fig-plr}
\end{figure}
The performance of CDA and EM-CDA is investigated as a function of the ground-truth packet loss rate in Fig. \ref{fig-plr}.
It is shown that the detection probability of the CDA schemes decreases as the packet loss rate increases, while the false alarm probability increases. EM-CDA is seen to be able to compensate for some of this performance loss, especially when using TE.

\begin{figure}[htp]
    \centering
    \subfigure[]{
    \includegraphics[scale=0.3]{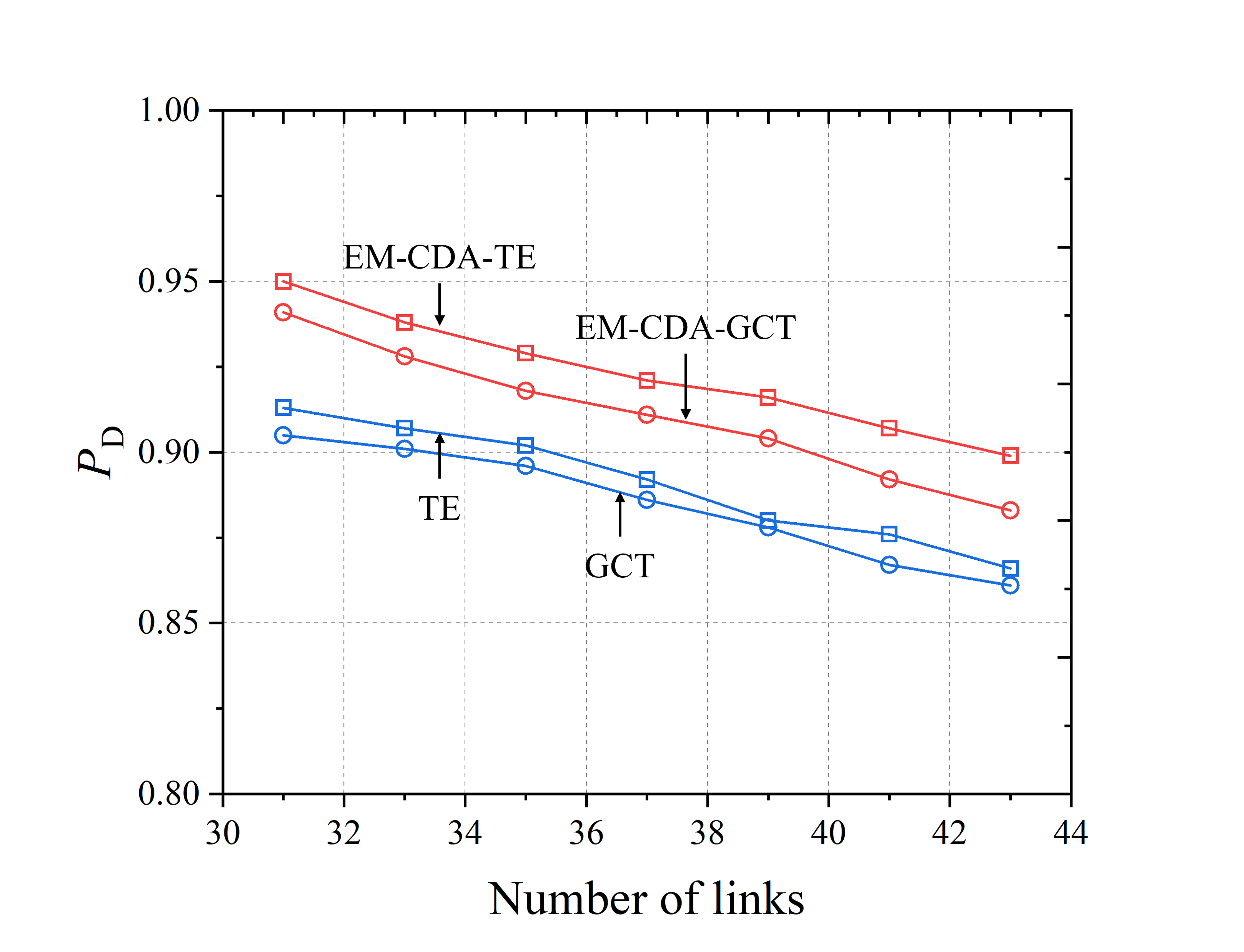}
    }
    \subfigure[]{
    \includegraphics[scale=0.3]{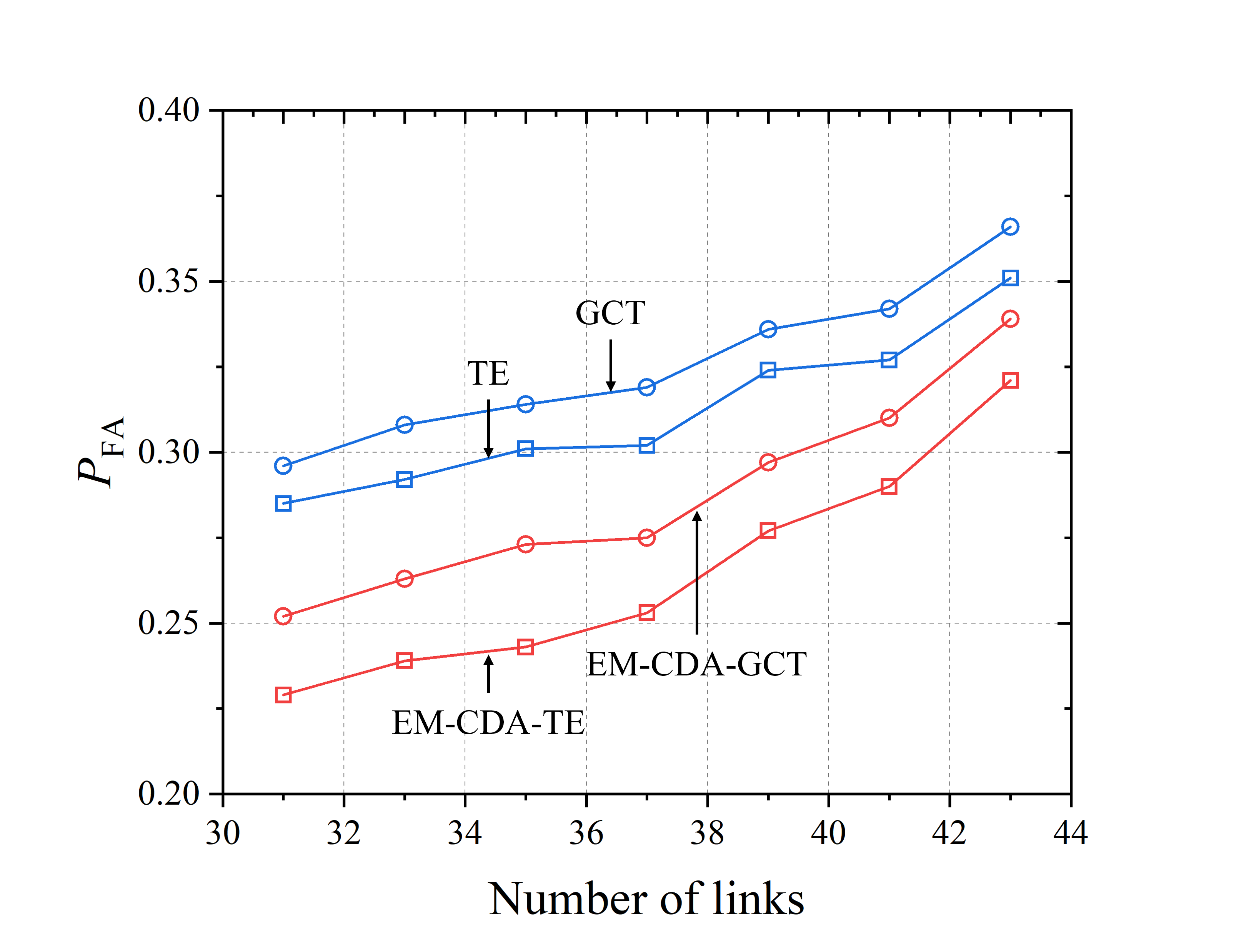}
    }
    \caption{(a) Probability of detection $p_{\rm D}$ and (b) probability of false alarm $P_{\rm FA}$ for topology inference versus the number of links for CDA and EM-CDA with GCT and TE causality metrics.}
    \label{fig-link}
\end{figure}
\begin{figure}[htp]
    \centering
    \subfigure[]{
    \includegraphics[scale=0.3]{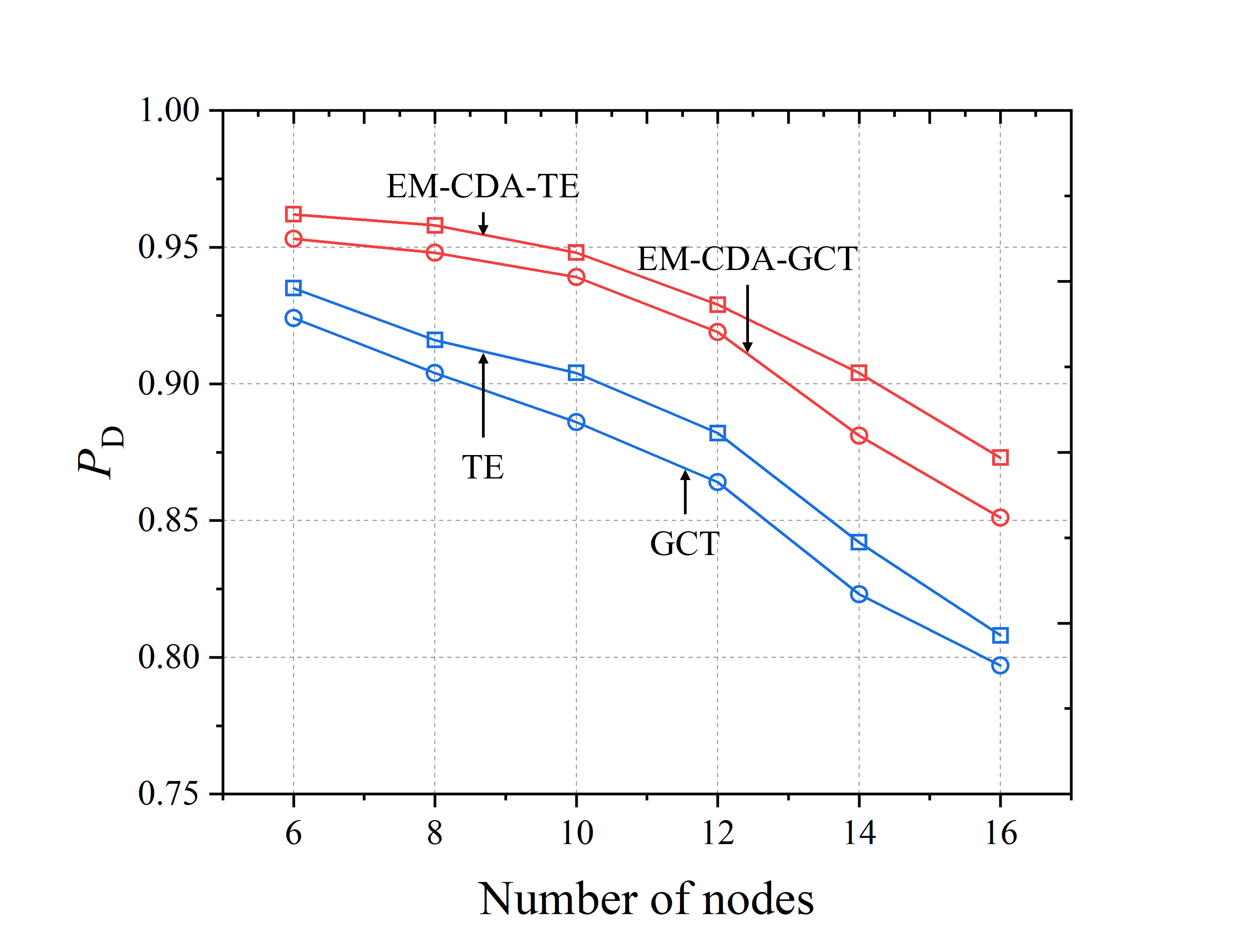}
    }
    \subfigure[]{
    \includegraphics[scale=0.3]{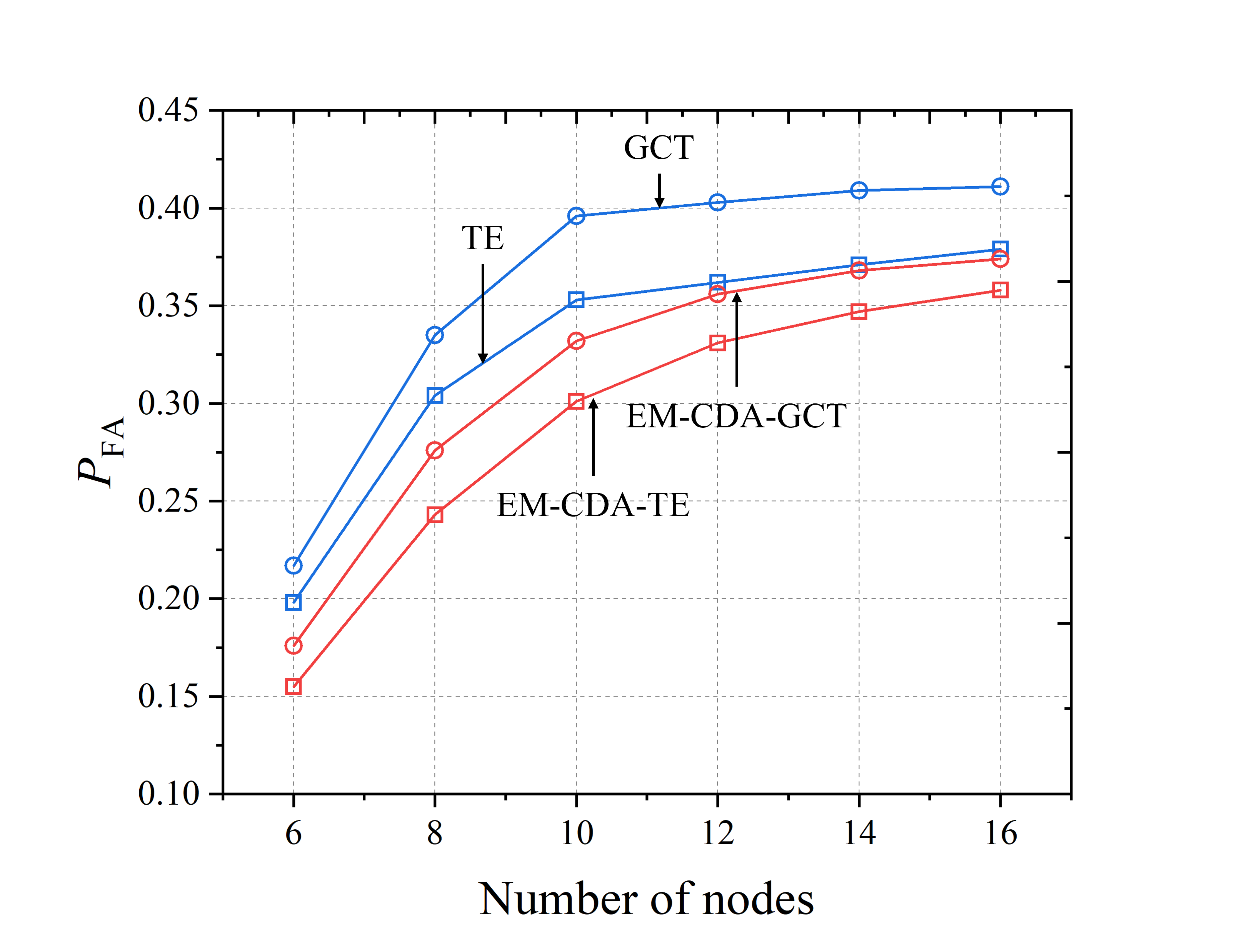}
    }
    \caption{(a) Probability of detection $p_{\rm D}$ and (b) probability of false alarm $P_{\rm FA}$ for topology inference versus the number of nodes for CDA and EM-CDA with GCT and TE causality metrics.}
    \label{fig-node}
\end{figure}
The relation between inference performance and the active link is investigated in Fig. \ref{fig-link}, while the number of nodes is fixed as $N=10$, and the number of the active link is changed from 31 to 43. With an increase in the active link, the mutual interference between nodes gets larger, but EM-CDA is able to retain its performance advantage as compared to CDA method. A similar conclusion is reached from Fig. \ref{fig-node}, which varies the number of nodes $N$ for a fixed fraction, 0.3, of active links.

\section{Conclusion} \label{Conclusion}
In this paper, we have introduced EM-CDA, a novel algorithm for passive network topology inference based on the observation of timing meta-data. The approach builds on the state-of-the-art causality discovery algorithm (CDA), and it addresses the important open problem of mitigating the effect of packet losses. Packet losses cause some of the timings of data packets to have no ACK packet counterparts, making CDA schemes potentially ineffective.  EM-CDA formulates the topology inference problem as the discrete maximum likelihood (ML) problem of identifying active links in the presence of latent packet losses. It alternates between estimation of packet losses and application of a CDA strategy. Numerical results based on NS-3 simulations of real-world networks show that EM-CDA outperforms CDA in terms of detection probability and false alarm probability by a range of 4\% to 12\% under a variety of network conditions accounting for different packet loss rates, number of nodes, and active links. Future work may investigate more accurate approximations of the EM algorithm, e.g., in the evaluation of the posterior distribution in the E step, as well as the adoption of a more detailed model to define the ML problem.


%



\bibliography{ref}

\end{document}